\documentclass[12pt, final]{article}

\usepackage{virgil_env} 
\usepackage{multirow}
\usepackage{fullpage}
\usepackage{nips10submit_e}
\usepackage{afterpage}
\usepackage{cite}
\usepackage{caption}
\nipsfinalcopy

\usepackage[utf8]{inputenc}
\usepackage[russian,english]{babel}
\usepackage{CJKutf8}

\graphicspath{{./}{figs/}{/Users/virgil/Documents/darkweb/analysis/figs/}}
\usepackage{booktabs,multirow,nicefrac}

\usepackage{xspace}
\usepackage{float}

\usepackage{tabu}

\newcommand{\dashrule}[1][black]{%
  \color{#1}\rule[\dimexpr.5ex-.2pt]{4pt}{.4pt}\xleaders\hbox{\rule{4pt}{0pt}\rule[\dimexpr.5ex-.2pt]{4pt}{.4pt}}\hfill\kern0pt%
}
\newcommand{\rulecolor}[1]{%
  \def\CT@arc@{\color{#1}}%
}

\usepackage{authblk}
\author[1]{Virgil Griffith}
\author[1]{Yang Xu}
\author[2]{Carlo Ratti}
\affil[1]{Urban Mobility Laboratory, Singapore-MIT Alliance for Research and Technology, Singapore, 138602}
\affil[2]{SENSEable City Laboratory, Massachusetts Institute of Technology, Cambridge, MA 02139, USA}

\usepackage{wasysym}

\usepackage{mathenv}
\usepackage{wasysym}

\title{Graph Theoretic Properties of the Darkweb}

\newcommand{\numnodes}{\ensuremath{7,178}\xspace}
\newcommand{\numedges}{\ensuremath{25,104}\xspace}
\newcommand{\COREsize}{\ensuremath{297}\xspace}
\newcommand{\muSPLwww}{\ensuremath{4.27}\xspace}
\newcommand{\muSPLdw}{\ensuremath{4.35}\xspace}
\newcommand{\muSPLcore}{\ensuremath{3.97}\xspace}
\newcommand{\percentzeroOD}{\ensuremath{87\%}\xspace}

\begin{document}

\maketitle

{\centering {\sffamily \today} \vskip10pt }


\begin{abstract}
We collect and analyze the darkweb (a.k.a. the ``onionweb'') hyperlink graph.  We find properties highly dissimilar to the well-studied world wide web hyperlink graph; for example, our analysis finds that $>\!\percentzeroOD$ of darkweb sites \emph{never} link to another site.   We compare our results to prior work on world-wide-web and speculate about reasons for their differences.  We conclude that in the term ``darkweb'', the word ``web'' is a connectivity misnomer.  Instead, it is more accurate to view the darkweb as a set of largely isolated dark silos.
\end{abstract}

\section{Introduction}
Graph theory has long been a favored tool for analyzing social relationships \cite{borgatti1998network} as well as quantifying engineering properties such as searchability \cite{wu2006query}.  For both reasons, there have been numerous graph-theoretic analyses of the World Wide Web (www) from the seminal \cite{barabasi1999emergence, barabasi2000scale, adamic2000power, kumar2000web, albert2001physics} to the modern \cite{meusel2015graph}.  Motivated by curiosity, we repeat these analyses for the notorious ``darkweb''.  The darkweb is sometimes loosely defined as ``anything seedy on the Internet'', but we define the darkweb strictly, as simply all domains underneath the ``.onion'' psuedo-top-level-domain\cite{rfc7686}, i.e., we define the darkweb to be synonymous with the onionweb.

The darkweb is infamously mysterious, and any insight into it for both harnessing it or informing social policy is welcome.  As far as we know we are among\cite{everton2012disrupting,darknet} the first to analyse the darkweb through graph theory.  We analyze the darkweb foremost because it's an interesting unexplored dataset, and second because, on the face of it, the world-wide-web and the darkweb are immensely similiar---both are browsed through a standard web-browser.  Therefore any differences between the structure of the darkweb versus the www likely indicate something about the societies inhabiting each.  For comparable work on the world-wide-web, see especially \cite{Serrano2007} and \cite{lehmberg2014graph}.

Newcomers are often confused about the properties of the onionweb.  Put simply, the onionweb is subset of the web where websites are identified not by a human-readable hostname (e.g., yahoo.com) or by a IP number (e.g., 206.190.36.45), but by a randomly generated 16-character address (specifically, a hash fingerprint).  Each website can be accessed via its hash, but it is very difficult to learn the IP number of a website from its hash.  The ``dark'' part means that, even though the website is accessible (via its hash), its IP number remains hidden.  Without an IP number, it is exceedingly difficult to trace the geographical origin of a communication.

\section{Data Collection}
Crawling the darkweb is not much harder than crawling the regular web.  In our case, we crawled the darkweb through the popular tor2web proxy onion.link.  Onion.link does all of the interfacing with Tor, and one can crawl all darkweb pages not requiring a login simply by setting a standard crawler to specifically crawl the domain \texttt{*.onion.link}.  Darkweb pages are written in the same HTML language as the regular web, and we crawled onion.link using the commercial service scrapinghub.com.  Starting from two popular lists of darkweb sites,\footnote{\texttt{http://directoryvi6plzm.onion} and \texttt{https://ahmia.fi/onions/}} we accessed each page and crawled all linked pages using breadth-first search.

Most graph-theoretic analyses of the www consider the page-level, i.e., each node is an individual URL.  The page-level description is important for engineering issues like crawlability, and one could certainly do a page-level analysis of the darkweb, however, it's unclear how informative this level of description is on social behavior---the page-level graph is influenced more by the various choices of content management system than anything social.

Because of this, we follow \cite{lehmberg2014graph} to \emph{aggregate by second-level domain} (for the onionweb the second-level domain is equivalent to \cite{lehmberg2014graph}'s ``pay-level domain'').  This means that links within a second-level domain are ignored as socially irrelevant self-connections.  In our darkweb graph, each vertex is a \emph{domain} and every directed edge from $u \rightarrow v$ means there exists a page within domain $u$ linking to a page within domain $v$.  The weight of the edge from $u \rightarrow v$ is the number of pages on domain $u$ linking to pages on domain $v$.

The Tor Project Inc., the creators and custodians of the darkweb, maintain basic darkweb statistics\cite{tormetrics}.  According to them, there are $\sim \!\! 60,000$ distinct, active .onion addresses. However, in our analysis we found merely \numnodes active .onion domains.  We attribute this high-discrepancy to various messaging services---particularly TorChat \cite{wiki:torchat}, Tor Messenger \cite{wiki:tormessenger}, and Ricochet \cite{wiki:ricochet}.  In all of these services, each user is identified by a unique .onion domain.

The darkweb has a notoriously high attrition rate---sites regularly appear and disappear.  This creates a substantial confound as there will be links to .onion domains that no longer exist.  To account for this, we only include sites which responded.  I.e., if we discover a link to a page on domain $v$, but domain $v$ could not be reached after $>\!10$ attempts across November 2016--February 2017, we delete node $v$ and  all edges to node $v$.  In our analysis, before pruning nonresponding domains we found a graph of 13,117 nodes and 39,283 edges.  After pruning, we have a graph of \numnodes nodes and \numedges edges (55\% and 64\% respectively).  In all results, we refer to this graph pruned of nonresponding domains as simply ``the darkweb graph''.

\section{Graph-theoretic Results}
The darkweb graph consists of exactly one weakly connected component (WCC)\footnote{A Weakly Connected Component is defined as a subset of a graph where every node in the subset, ignoring edge direction, can reach every other node in the subset.  Finding \emph{exactly one} weakly connected component is entailed by our original list of seed domains itself being part of the darkweb.} of \numnodes nodes and \numedges edges.

\begin{figure}[h!bt]
	\centering
	\subfloat[darkweb in-degree]{ \includegraphics[width=0.5\textwidth]{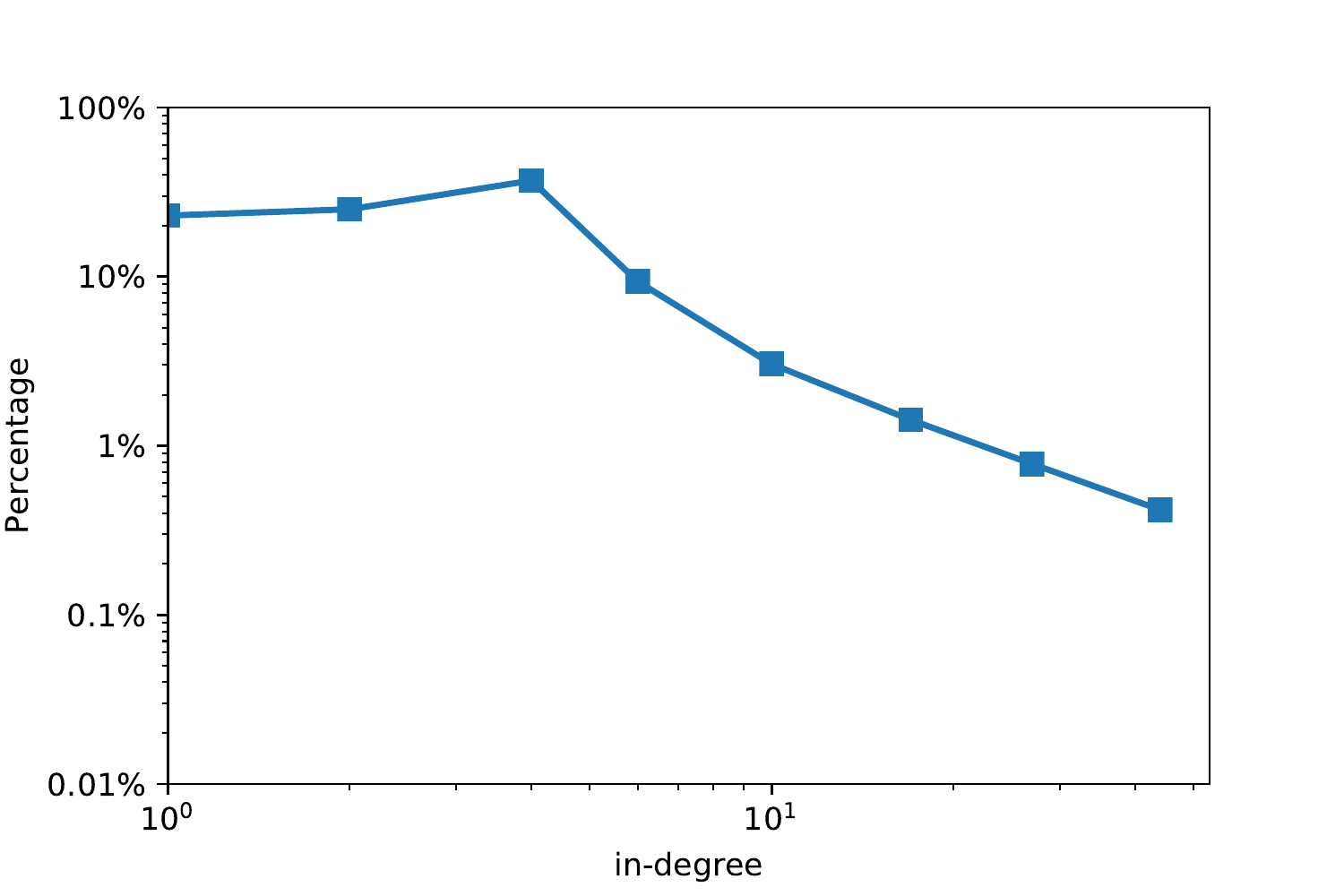} \label{fig:dw_id} }
	\subfloat[darkweb out-degree]{ \includegraphics[width=0.5\textwidth]{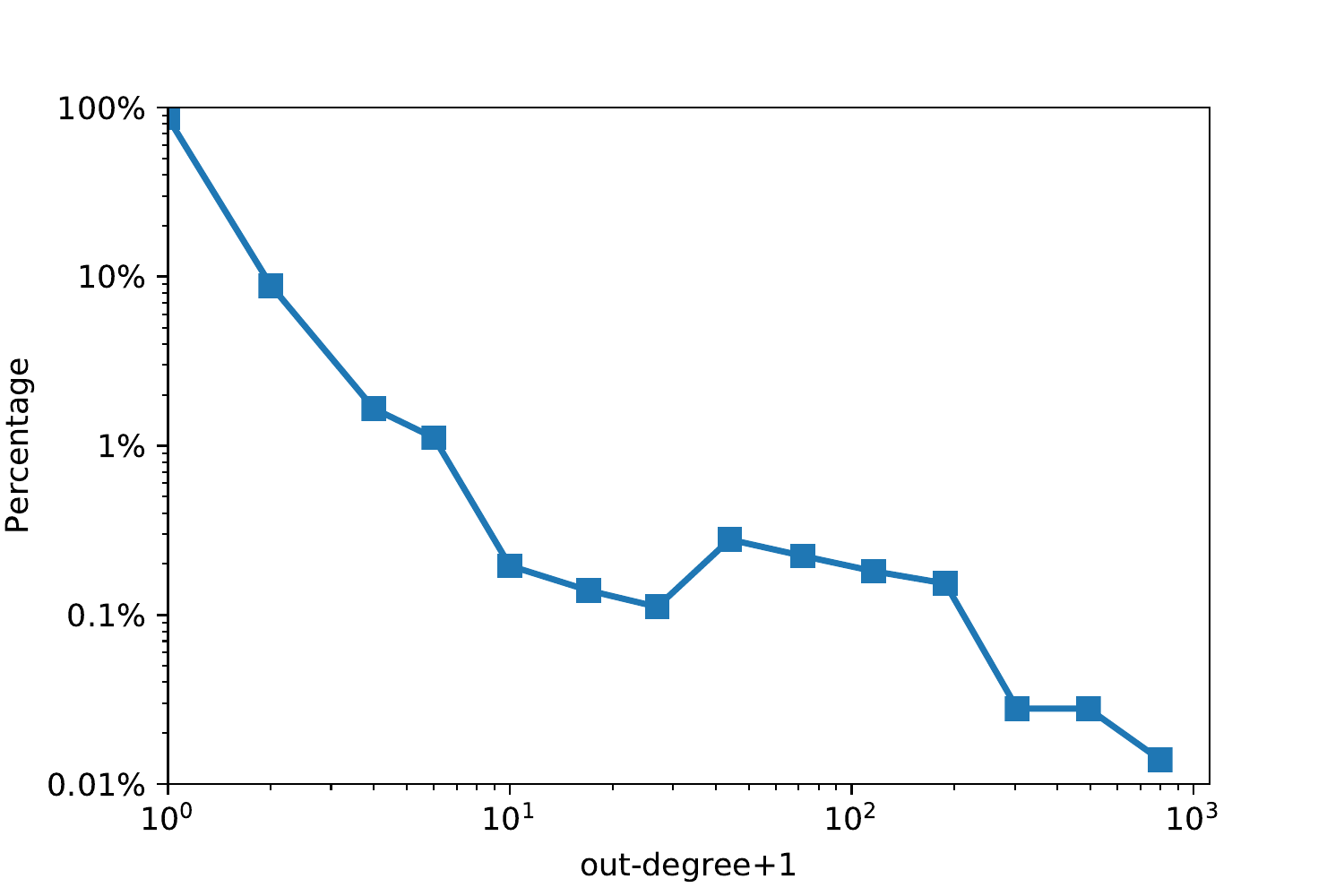} \label{fig:dw_od} }
    \\
	\subfloat[darkweb pagerank]{ \includegraphics[width=0.5\textwidth]{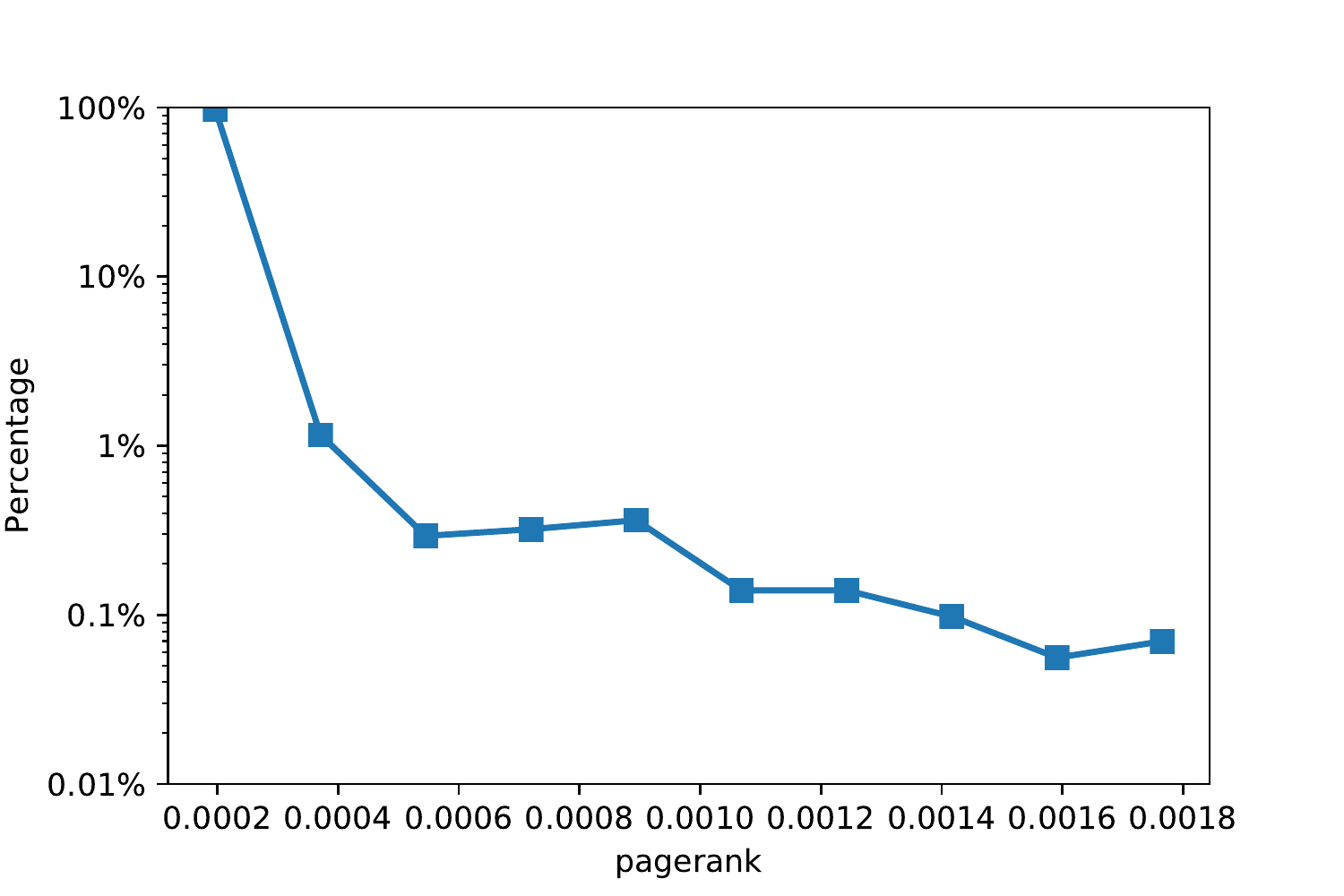} \label{fig:dw_pr} }
	\subfloat[darkweb edgeweights]{ \includegraphics[width=0.5\textwidth]{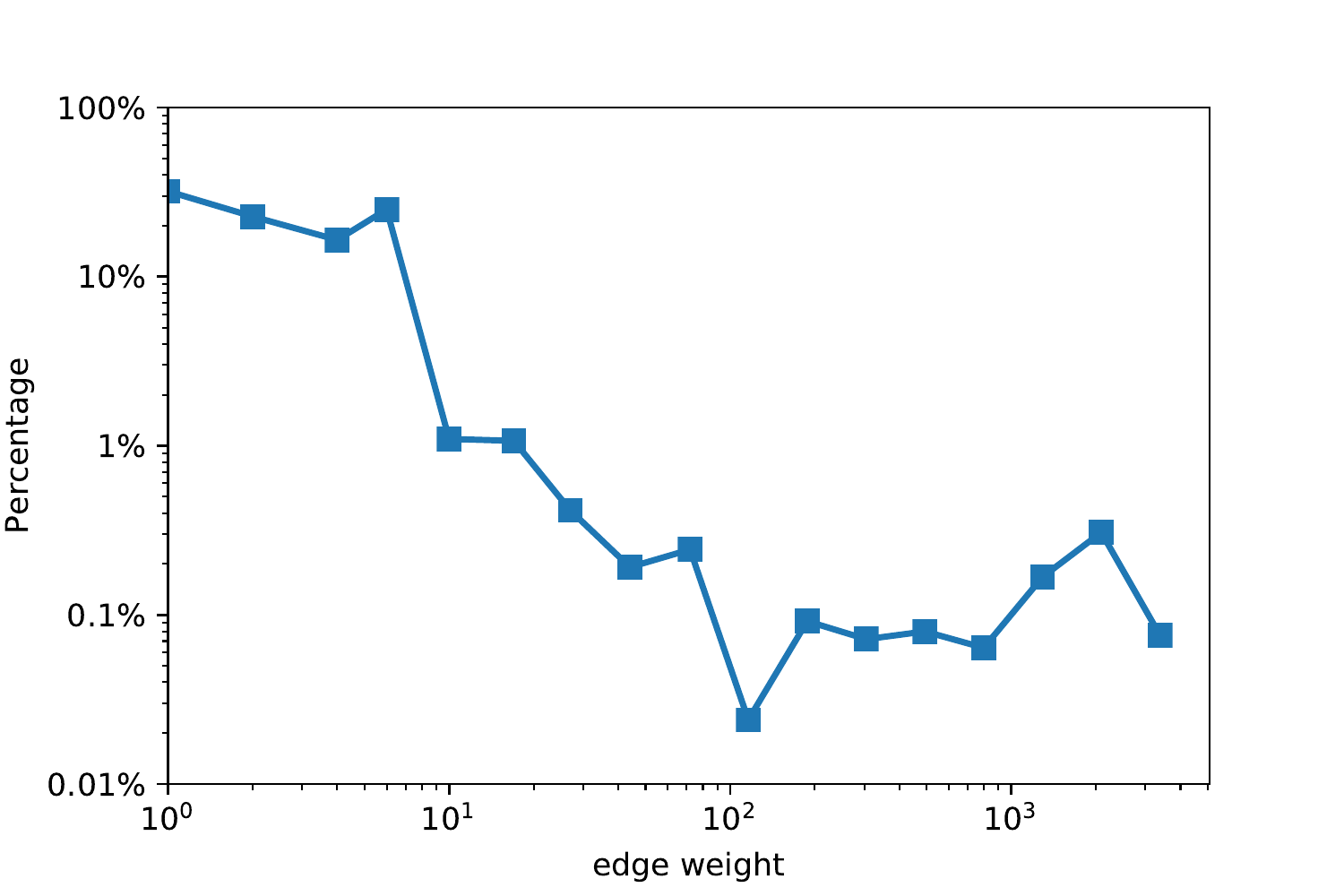} \label{fig:dw_ew} }	
	\caption{The distribution of the in-degree, out-degree, pagerank, and edgeweights. In \figref{fig:dw_pr} we exclude the three domains with the highest pagerank (they are listed in \tblref{tbl:topN} in the Appendix) because they are such extreme outliers.  For all plots with a log-scale axis, we follow following \cite{lehmberg2014graph, meusel2015graph} to use the Fibonacci binning from \cite{vigna13fibonacci}.}
	\label{fig:degree_and_pr}
\end{figure}

The first step is to quanitatively analyze the degree distributions, which we do in \figref{fig:degree_and_pr}.  Figures \ref{fig:dw_id}, \ref{fig:dw_od}, and \ref{fig:dw_ew} vaguely resemble a powerlaw, but the \texttt{plfit} tool using the methods from \cite{clauset2009power} report there's not enough orders of magnitude in the data to affirm or deny a powerlaw.  In \figref{fig:dw_id} we see that $\sim\! 30\%$ of domains have \emph{exactly one} incoming link---of which $62\%$ come from one of the five largest out-degree hubs.  Intrigued by the impact of these five large out-degree hubs, we found that almost all nodes ($78\%$) received a connection from at least one of them.  For those curious about these specifics of these high out-degree hubs, the top fifteen hubs are listed in the Appendix (Table \ref{tbl:tophubs}).  In \figref{fig:dw_od} we see our most striking feature---that fully \percentzeroOD of sites \emph{do not link to any other site}.  This sole feature has immense impact on all graph-theoretic measures.  In \figref{fig:dw_pr} we see that $>\!98\%$ of domains are tied for the lowest pagerank \cite{pagerank} of $2.095\textsc{e}\textnormal{-}4$.  In \figref{fig:dw_ew}, we seen that when a site does link another, $32\%$ of the time it's only a single page linking out.  All together, the onionweb is a sparse hub-and-spoke place.

\begin{figure}[hbt]
\centering

\hbox{\centering \hspace{11em} \includegraphics[width=0.6\textwidth]{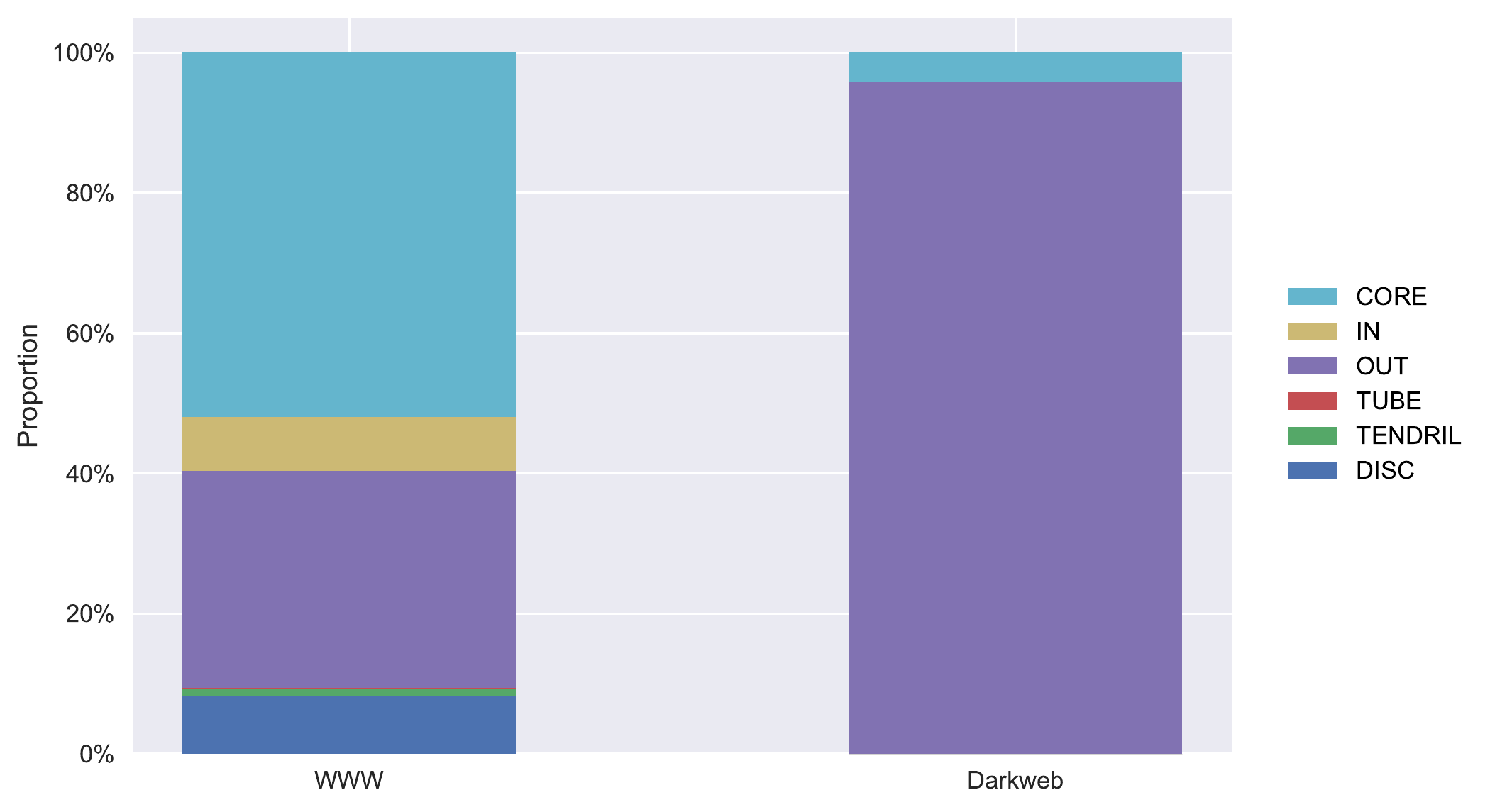} }

\vspace{1.5em}

\begin{tabular}{l | r r r r} \toprule
& \multicolumn{2}{c}{World-Wide-Web \cite{lehmberg2014graph}} & \multicolumn{2}{c}{Darkweb} \\
\midrule
	CORE         & 22.3M & 51.94\% & 297 & 4.14\% \\
	IN           & 3.3M  &  7.65\% & 0 & 0.0\% \\
	OUT          & 13.3M & 30.98\% & 6881 & 95.86\% \\
	TUBES        & 17k   &   .04\% & 0 & 0.0\% \\
	TENDRILS     & 514k  &   1.2\% & 0 & 0.0\% \\
	DISCONNECTED & 3.5M  &   8.2\% & 0 & 0.0\% \\
\bottomrule
\end{tabular}

\caption{Bow-tie decomposition comparing the www \cite{lehmberg2014graph} versus the darkweb.}
\label{fig:bowtie}

\end{figure}

Our second result is a \emph{bow-tie decomposition} from \cite{broder2000graph}, shown in \figref{fig:bowtie}.  A bowtie decomposition divides the nodes of a directed graph into six disjoint parts, they are:
\begin{enumerate}
    \item \textsc{CORE} --- Also called the ``Largest Strongly Connected Component''.  It is defined as the largest subset of nodes such that there exists a directed path (both directions---from $u \rightarrow \cdots \rightarrow v$ as well as $v \rightarrow \cdots \rightarrow u$) between every pair of nodes in the set.
    \item \textsc{IN} --- The set of nodes, excluding those in the \textsc{CORE}, that are ancestors of a \textsc{CORE} node.
    \item \textsc{OUT} --- The set of nodes, excluding those in the CORE, that are descendants of a \textsc{CORE} node.
    \item \textsc{TUBES} --- The set of nodes, excluding those in the \textsc{CORE}, \textsc{IN}, and \textsc{OUT}, who have an ancestor in \textsc{IN} as well as a descendant in \textsc{OUT}.
    \item \textsc{TENDRILS} --- Nodes that have an ancestor in \textsc{IN} but do not have a descendant in \textsc{OUT}.  Also, nodes that have a descendant in OUT but do not have an ancestor in \textsc{IN}.
    \item \textsc{DISCONNECTED} --- Everything else.
\end{enumerate}

We compare our results to the www results from \cite{Serrano2007, lehmberg2014graph, meusel2015graph}.  We chose these reference points due to the size of their crawls and the rigor of their analyses.  The most obvious difference between the world-web-web and the darkweb is that the darkweb only contains a \textsc{CORE} and an \textsc{OUT} component.  We attribute this to the extraordinarily low out-degrees from \figref{fig:dw_od}.  For the curious, the top 50 nodes of the \textsc{CORE} are provided in the Appendix (Table \ref{tbl:CORE}).

\begin{figure}[h!bt]
	\centering
	\subfloat[\hspace{1em}word wide web\newline $\mu=\muSPLwww$]{ \includegraphics[width=0.33\textwidth]{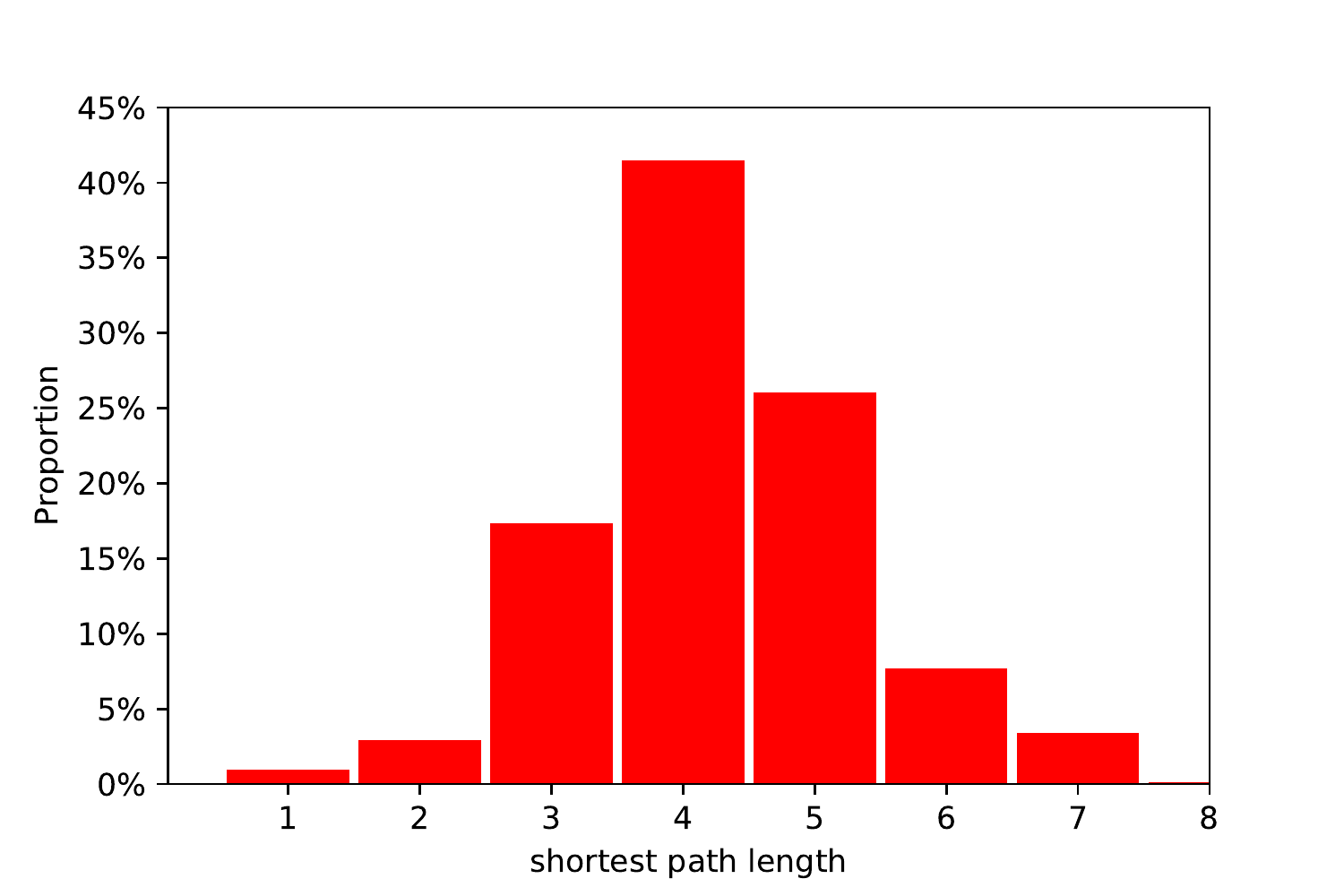} \label{fig:www_spl} }
	\subfloat[\hspace{2em}darkweb\newline $\mu=\muSPLdw$]{ \includegraphics[width=0.33\textwidth]{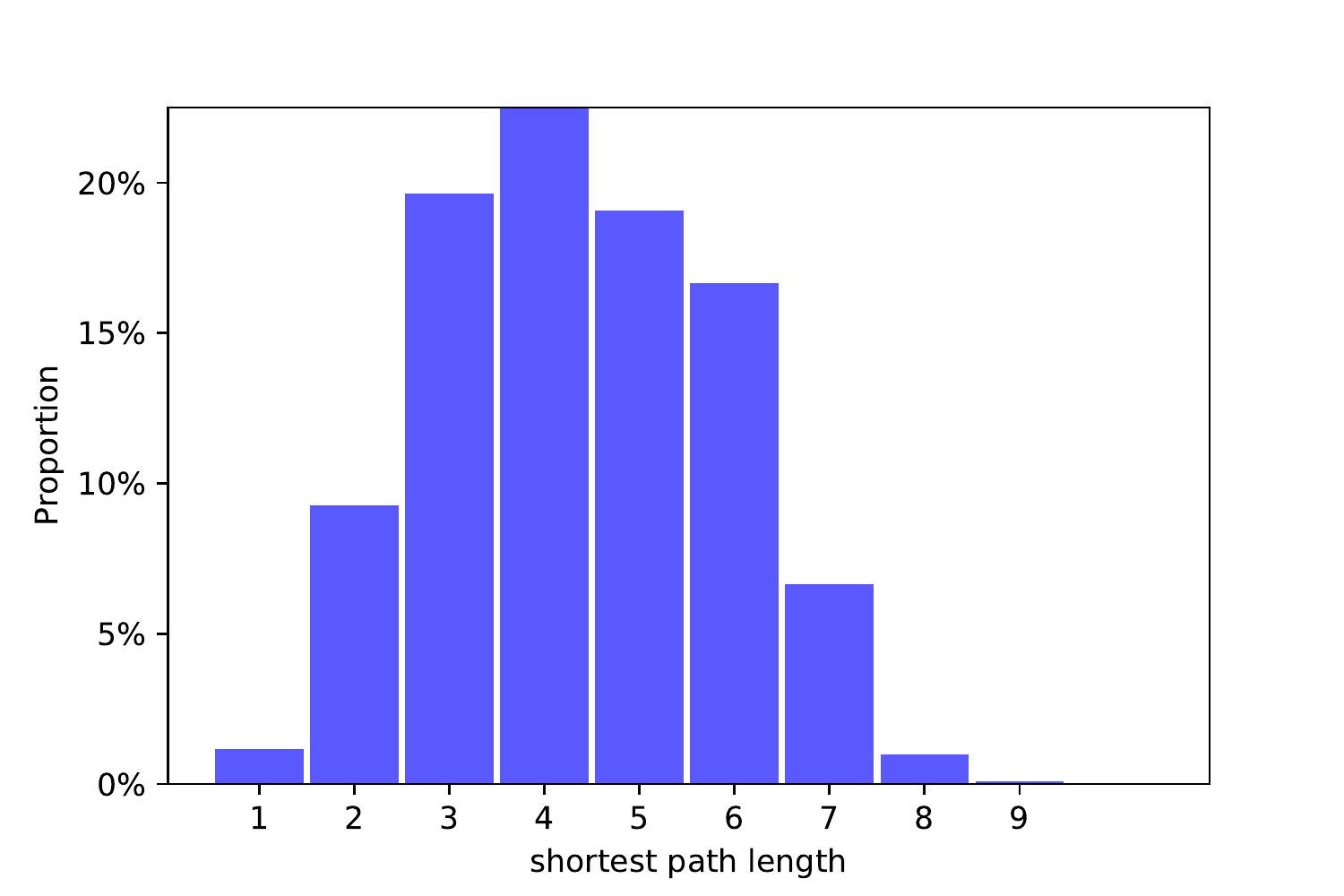} \label{fig:dw_spl} }
	\subfloat[\hspace{1em}darkweb \textsc{CORE}\newline $\mu=\muSPLcore$]{ \includegraphics[width=0.33\textwidth]{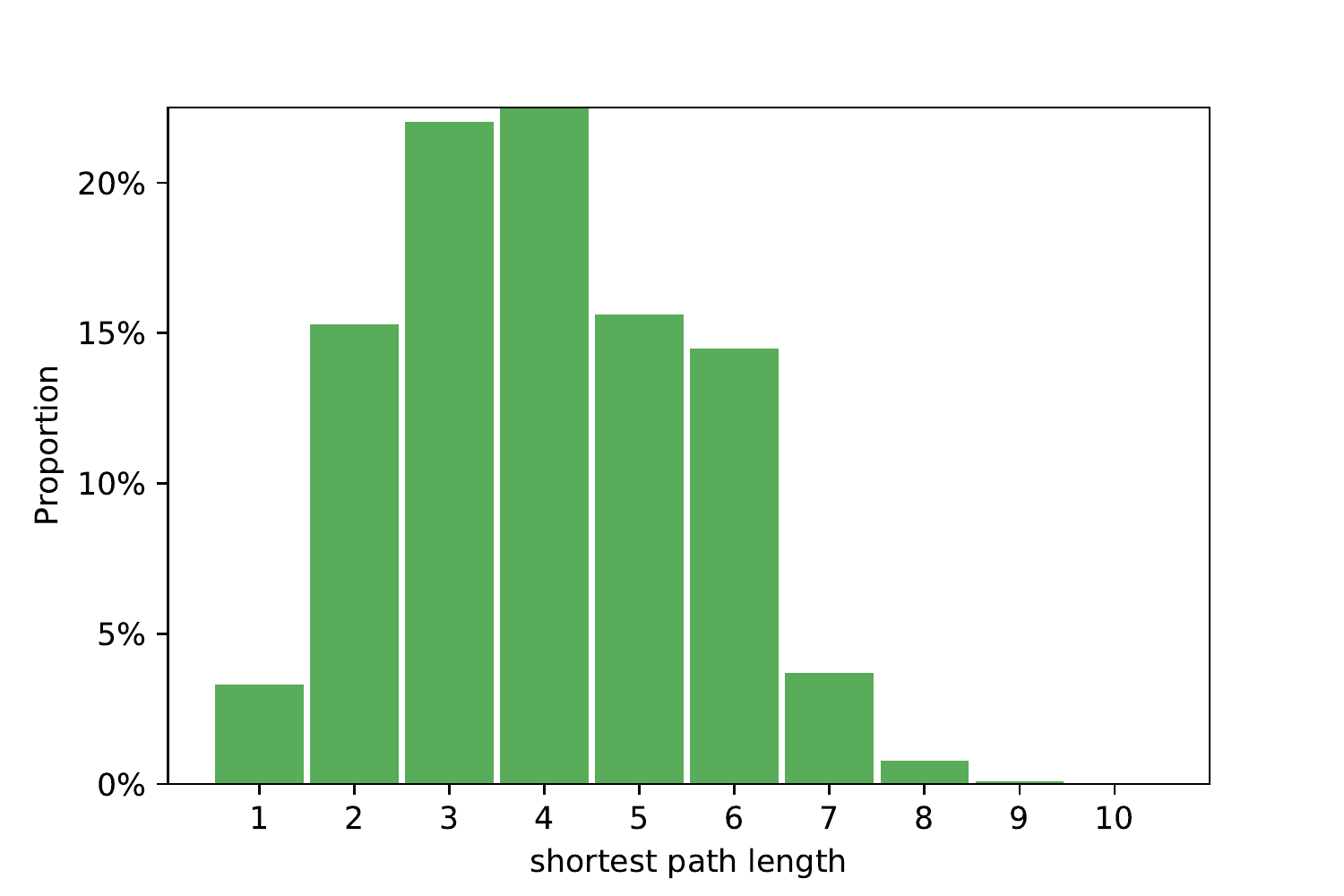} \label{fig:dw_core_spl} }	
	\caption{Comparing shortest path lengths between the world-wide-web and the darkweb considering directed edges.  Whereas in the www $56.65\%$ of node pairs have have $\infty$ path-length (no path connecting them), in the darkweb $91.89\%$ of node-pairs have no path connecting them.  Moreover, even within that $8.11\%$ of pairs with a directed path between them, the darkweb's average SPL ($\mu = \muSPLdw$) is \emph{higher} than that of the www ($\mu = \muSPLwww$).}
	\label{fig:spl}
\end{figure}

For our third result, we examine the darkweb's internal connectivity via shortest-path-length (SPL), shown in \figref{fig:spl}.  First, for all pairs of nodes $\{u,v\}$ in the darkweb, only 8.11\% are connected by a directed path from $u \rightarrow \cdots \rightarrow v$ or $v \rightarrow \cdots \rightarrow u$.  This is drastically lower than the $\sim\!\!43.42\%$ found in the www \cite{lehmberg2014graph}.  We again attribute this to the low out-degree per \figref{fig:dw_od}.  Of the connected pairs, the darkweb's average shortest path length is \muSPLdw compared to the \muSPLwww in the world-wide-web \cite{lehmberg2014graph}.  It's surprising to see a graph as small as the darkweb have a higher mean SPL than the entire world-wide-web, and is a testament to how sparse the darkweb graph really is.  In \figref{fig:dw_core_spl} we plot the distribution of SPLs for the \COREsize nodes of the \textsc{CORE}, to our surprise, the mean SPL within the \textsc{CORE} is \muSPLcore, only $9\%$ less than the entire darkweb.  From this we conclude the \textsc{CORE} is not any kind of densely interconnected core.

\subsection{Robustness and Fragility}
\begin{figure}[h!bt]
	\centering

	\subfloat[removing nodes by lowest in-degree]{ \includegraphics[width=0.45\textwidth]{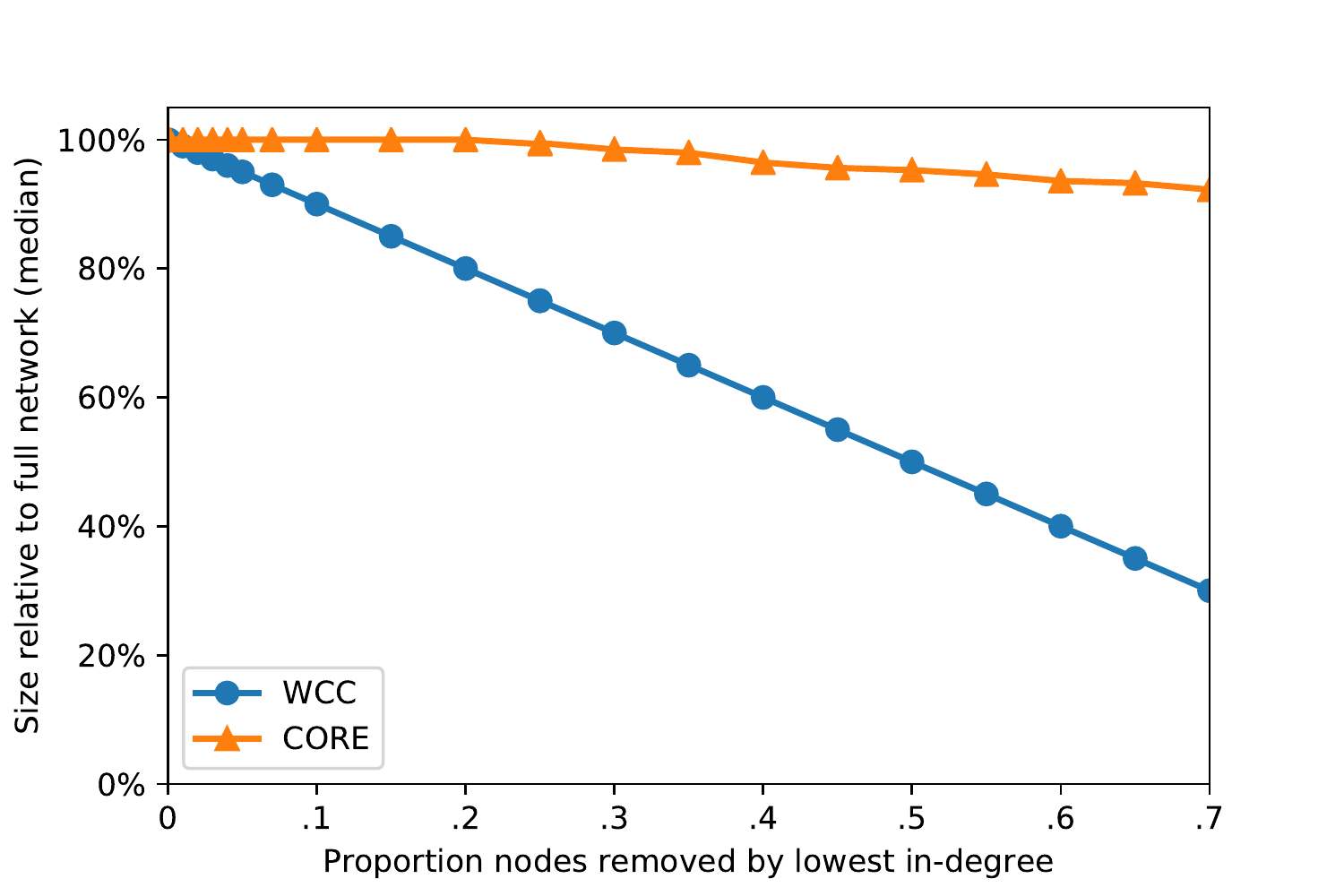} \label{fig:dw_removing_indgree_lowest} }
	\subfloat[removing nodes by highest in-degree]{ \includegraphics[width=0.45\textwidth]{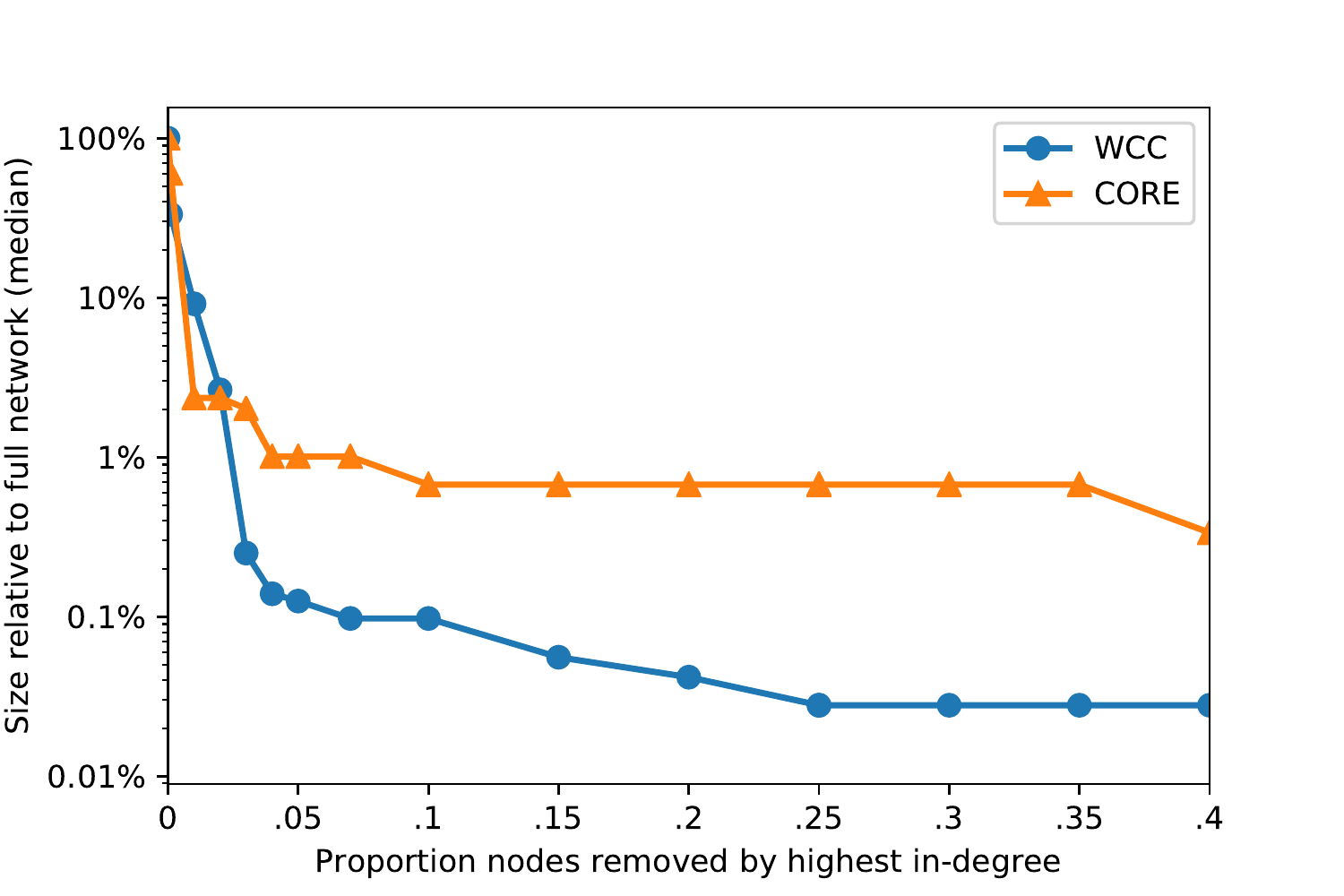} \label{fig:dw_removing_indgree_highest} }

	\subfloat[removing nodes by lowest harmonic centrality]{ \includegraphics[width=0.45\textwidth]{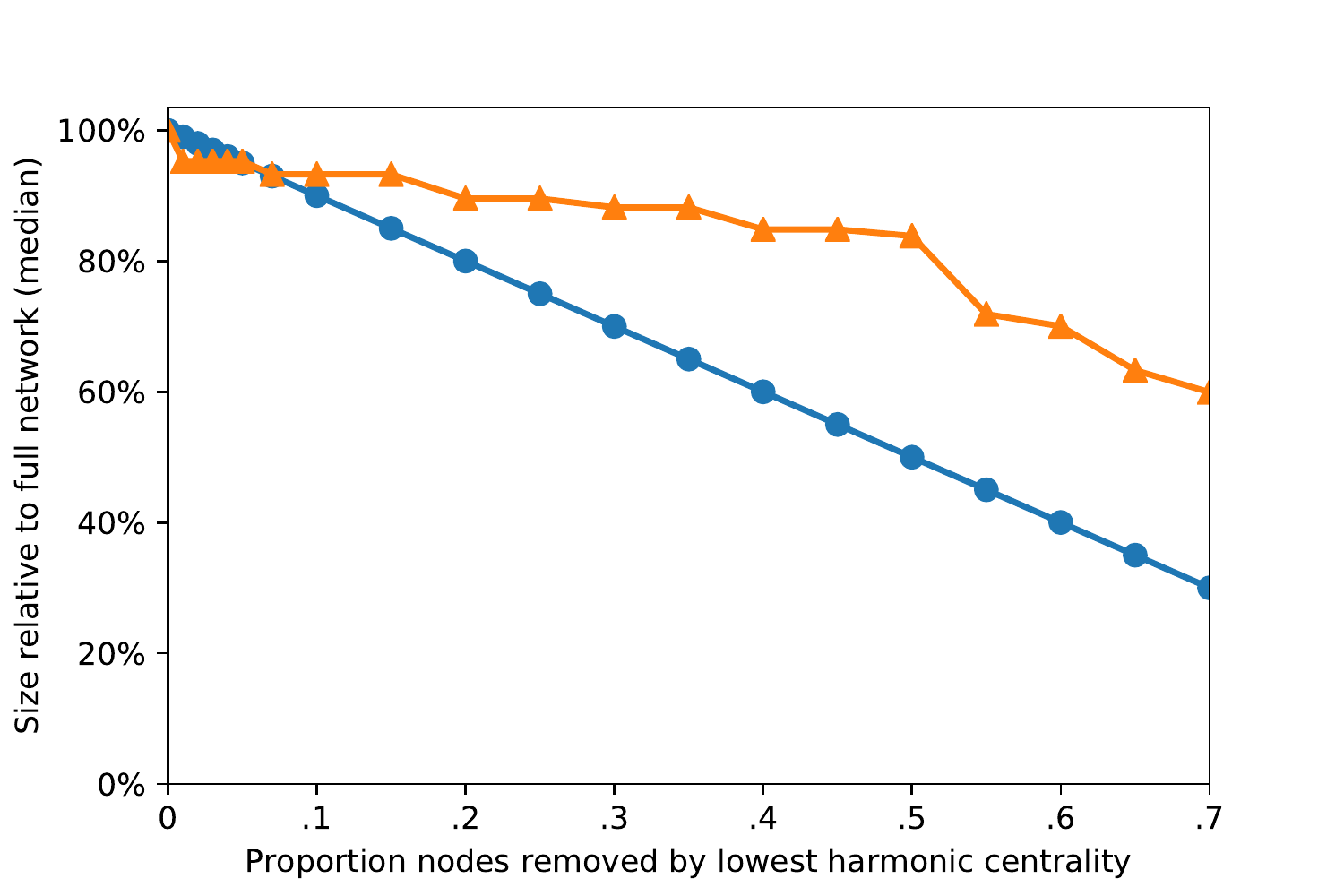} \label{fig:dw_removing_harmonic_lowest} }
	\subfloat[removing nodes by highest harmonic centrality]{ \includegraphics[width=0.45\textwidth]{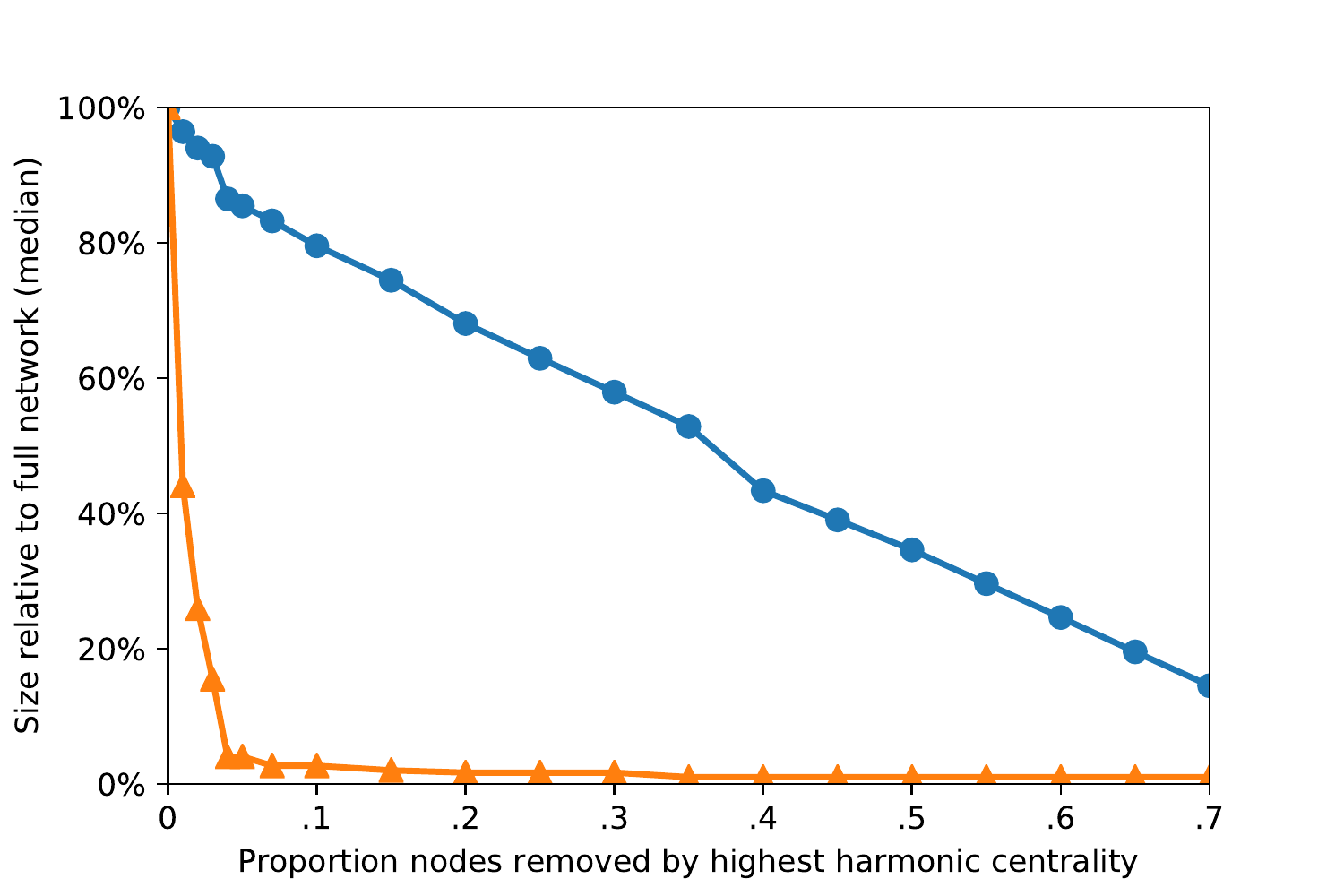} \label{fig:dw_removing_harmonic_highest} } \\

	\subfloat[removing nodes by lowest pagerank]{ \includegraphics[width=0.45\textwidth]{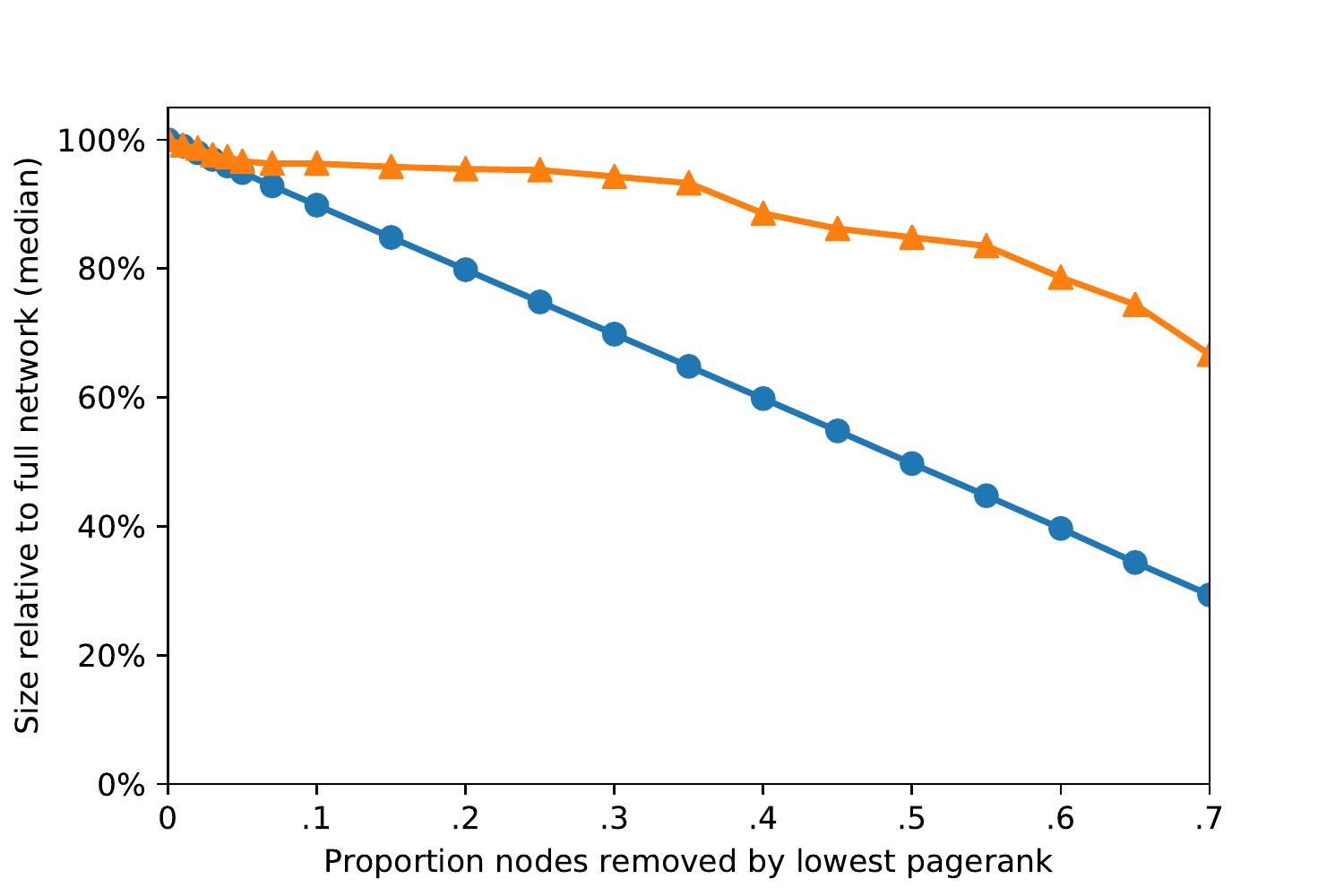} \label{fig:dw_removing_pagerank_lowest} }
	\subfloat[removing nodes by highest pagerank]{ \includegraphics[width=0.45\textwidth]{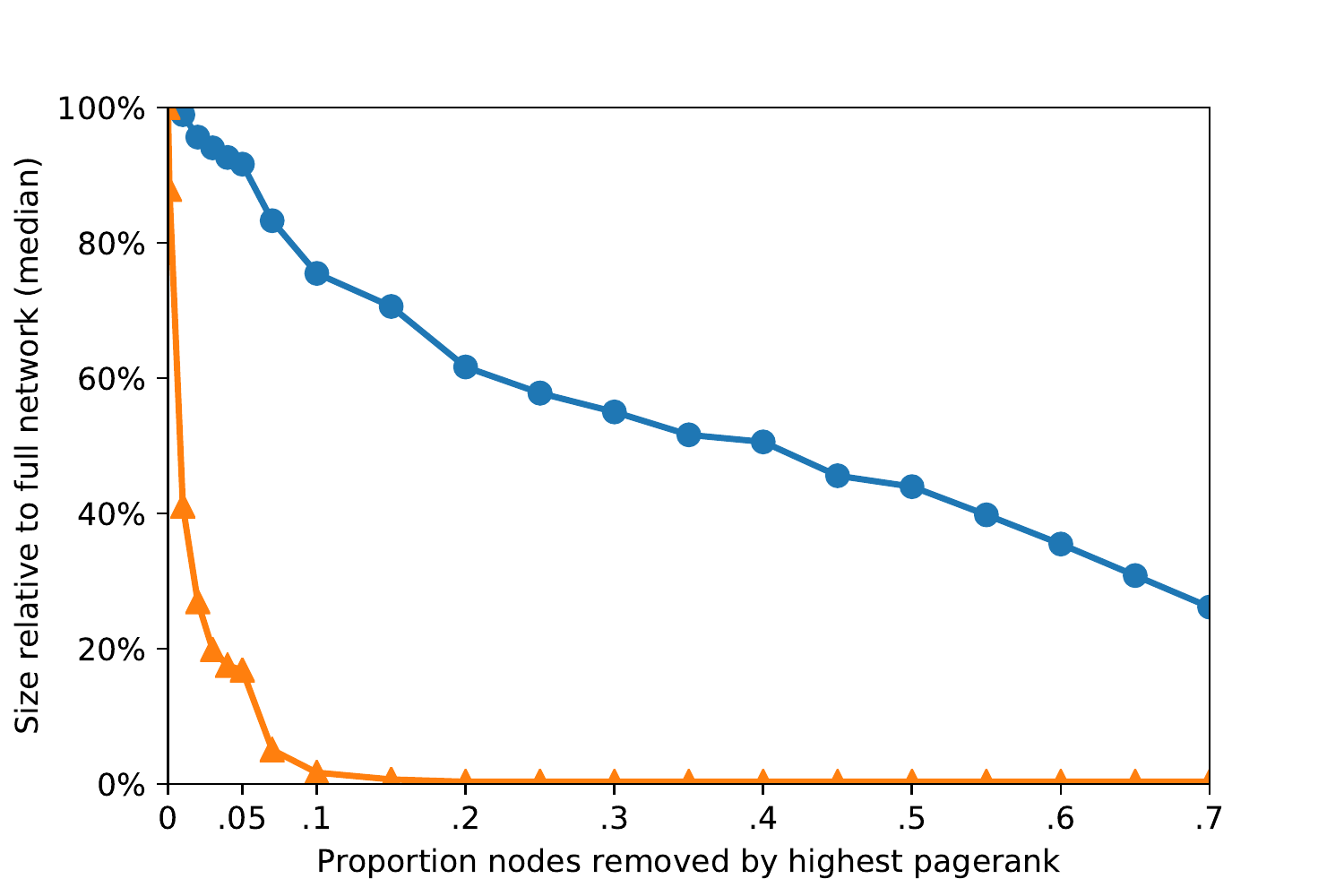} \label{fig:dw_removing_pagerank_highest} }

	\caption{Deleting nodes from the darkweb graph and seeing how quickly the WCC and \textsc{CORE} disintegrate.  In all plots, we shuffled the order of nodes with the same value until reaching stable statistics, e.g., in \figref{fig:dw_pr}, $98\%$ of nodes are tied for the lowest pagerank; so when removing only $10\%$ of the nodes (e.g., \figref{fig:dw_removing_pagerank_lowest}), it's ambiguous which nodes should be deleted first.  So in our analysis we shuffled the ordering of the nodes with the same value and recomputed sizes of the WCC/\textsc{CORE} until the median was stable.}
	\label{fig:knockouts}
\end{figure}

In \figref{fig:knockouts} we show our fourth result---how quickly the entire network (WCC) as well as the \textsc{CORE} disintegrates under node removal.  In Figures \ref{fig:dw_removing_indgree_lowest} and \ref{fig:dw_removing_indgree_highest} we see the familiar resistance to random failure yoked with fragility to targeted attacks in the spirit of \cite{bollobas2004robustness}.  In \figref{fig:dw_removing_indgree_highest}, we see that, unlike the www \cite{lehmberg2014graph}, the WCC is \emph{more susceptible to high in-degree deletions than the \textsc{CORE}}.  This elaborates the view from \figref{fig:dw_core_spl} that the \textsc{CORE} is, in addition to not being strongly interconnected, is also not any kind of high in-degree nexus.

Figures \ref{fig:dw_removing_harmonic_lowest} and \ref{fig:dw_removing_harmonic_highest} show the breakdown when removing central nodes.  In \figref{fig:dw_removing_harmonic_lowest} the CORE is largely unaffected by low centrality deletions.  In \figref{fig:dw_removing_harmonic_highest} we see that although the \textsc{CORE} isn't disproportionately held together by high in-degree nodes, it is dominated by very central nodes.

Comparing Figures \ref{fig:dw_removing_indgree_highest} and \ref{fig:dw_removing_pagerank_highest}, we see the \textsc{CORE} relative to the entire network consists of more high-pagerank nodes than high in-degree nodes.  This implies \textsc{CORE} nodes are not created by their high-indegree (\figref{fig:dw_removing_indgree_highest}), but by their high centrality, amply corroborated by Figures \ref{fig:dw_removing_harmonic_lowest} and \ref{fig:dw_removing_harmonic_highest}.  Likewise, \figref{fig:dw_removing_pagerank_lowest} recapitulates \figref{fig:dw_removing_indgree_lowest}, that nodes with especially low in-degree or centrality are, unsurprisingly, not in the \textsc{CORE}.

Given the hub-and-spoke nature of the graph, if the goal was \emph{destroy the connectivity of the hyperlink graph}, the most effective method would be to attack the central hubs---but which hubs?  In the Appendix (\figref{fig:more_robustness}) we see the WCC breaks down at roughly the same rate when removing either the high in-degree or high out-degree nodes.

\subsection{Reciprocal Connections}
\begin{figure}[h!bt]
	\centering
    \subfloat[With red outliers]{ \includegraphics[width=0.45\textwidth]{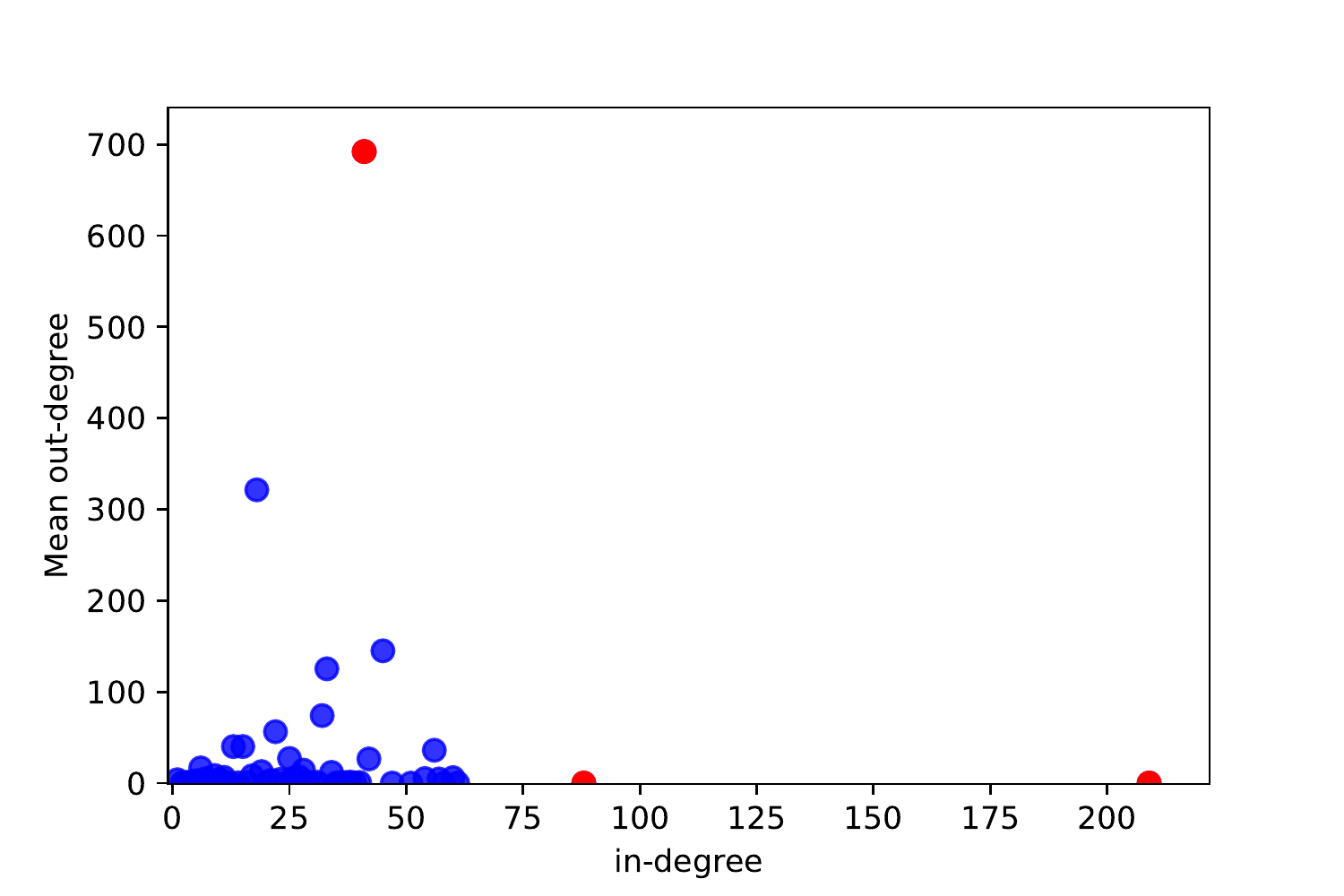} \label{fig:fig9_withoutliers} }
    \subfloat[Without red outliers]{ \includegraphics[width=0.45\textwidth]{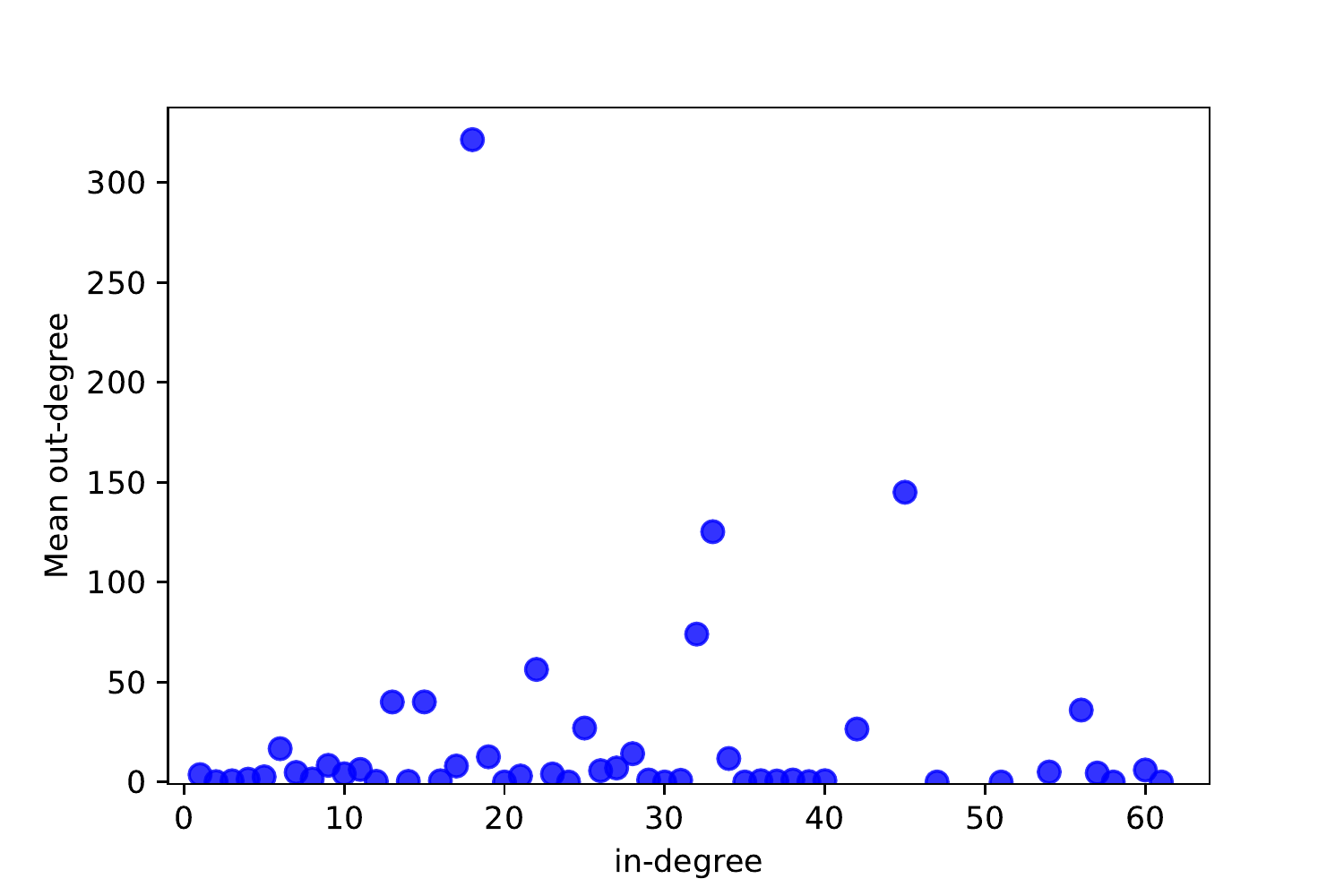} \label{fig:fig9_withoutoutliers} }    
	\caption{Average out-degree as a function of the in-degree.  \figref{fig:fig9_withoutoutliers} is the same as \ref{fig:fig9_withoutliers} but pruned of the three red outliers in \ref{fig:fig9_withoutliers}.}
	\label{fig:outdeg_as_func_of_indeg}
\end{figure}

In \cite{Serrano2007} they stress the importance of reciprocal connections in maintaining the www's graph properties.  We compute two of their measures.  First, we compute \cite{Serrano2007}'s measure $\frac{\langle k_{in}k_{out} \rangle }{ \langle k_{in} \rangle \langle k_{out} \rangle } = \frac{ \mathbb{E}[ k_{in} k_{out} ] }{ \mathbb{E}[ k_{in} ] \mathbb{E}[ k_{out} ] }$, to quantify in-degree and out-degree's deviation from independence.  For the darkweb, we arrive at $\frac{\langle k_{in}k_{out} \rangle}{\langle k_{in} \rangle \langle k_{out} \rangle}=3.70$.  This is in the middle of the road of prior estimates of the www, and means that the out-degree and in-degree are positively correlated.  To be a better view, we plot the average out-degree as a function of the in-degree, given as,
\begin{equation}
\langle k_{out}(k_{in}) \rangle = \frac{1}{N_{k_{in}}} \sum_{i \in \Upsilon(k_{in})} k_{out,i} \; ,
\label{eq:serrano}
\end{equation}
which is simply ``For all nodes of a given in-degree, what is the mean out-degree?''.  The results are given in \figref{fig:outdeg_as_func_of_indeg}; in short, in the darkweb there's no obvious pattern to the relationship between in-degree and out-degree, and is in fact mostly flat.

\section{Discussion}
\label{sect:discussion}

Casual inspection of \figref{fig:degree_and_pr} shows that the darkweb is indeed a very different graph than what in theworld wide web \cite{barabasi1999emergence, barabasi2000scale, adamic2000power, kumar2000web, albert2001physics, Serrano2007, meusel2015graph}---particularly, the glaring fact of so little linking to other websites (\figref{fig:dw_od}).
The fundamental question is why there's so little linking.  For this, the two major explanations are:
\begin{itemize}
    \item \textbf{The technological explanation.} In the darkweb, sites go up and go down all the time.  Why bother linking if there's little chance that the destination will still exist?
    \item \textbf{The social explanation.}  As-is, people who create sites on the darkweb are cut from a different social cloth than those who create sites on the www.
\end{itemize}

\begin{figure}[h!bt]
	\centering
    \includegraphics[width=0.7\textwidth]{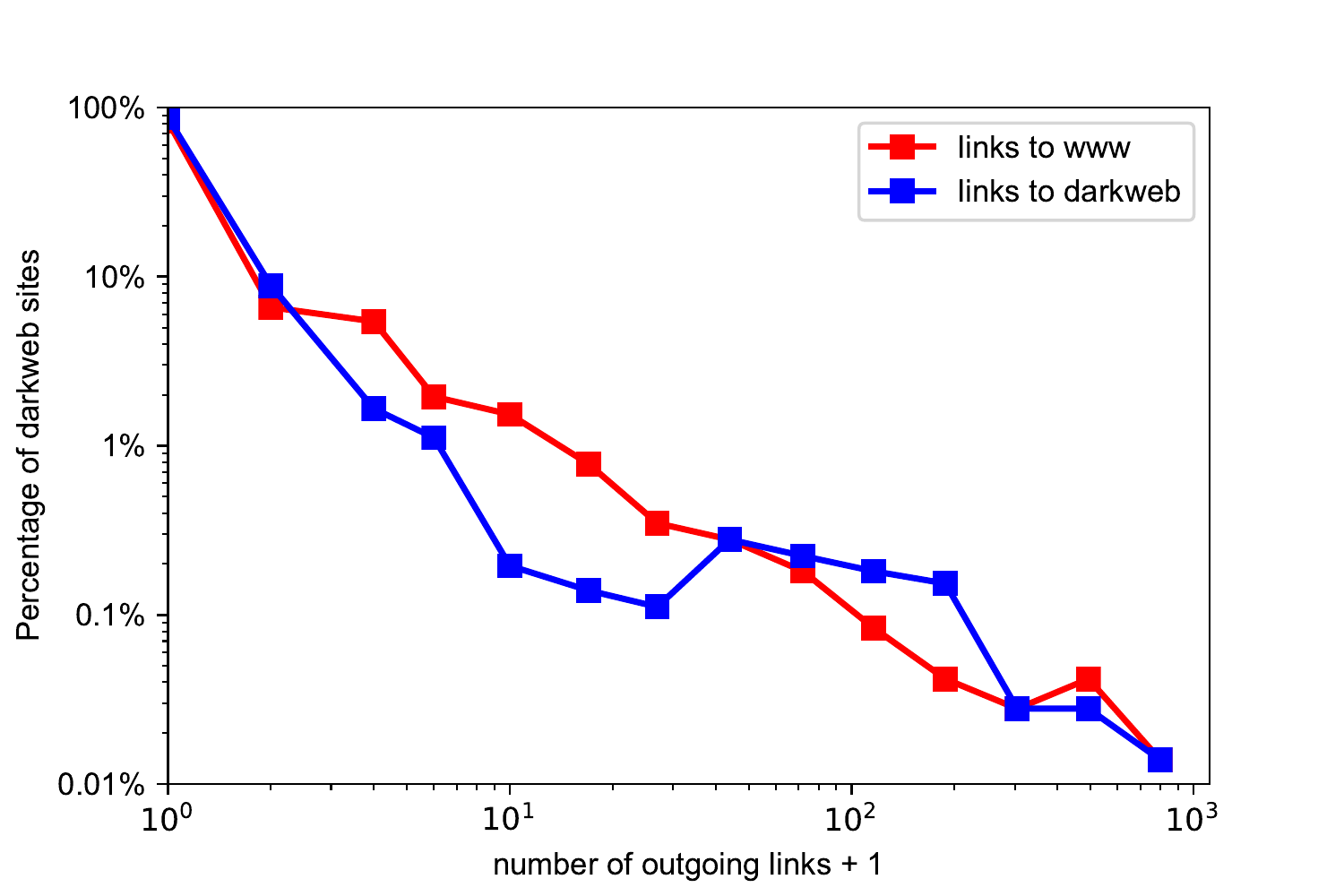} 
	\caption{Comparing the rates of the darkweb sites linking to the www versus linking to other darkweb sites, we see they are essentially the same.}
    \label{fig:dw2www}
\end{figure}

To disambiguate these two we performed a second crawl collecting instances of darkweb sites linking to the www and compared the relative rates of outbound linking, resulting in \figref{fig:dw2www}.  From the essentially equal rates of outbound linking to the www as well as the darkweb, we conclude:

\begin{enumerate}
    \item The low outbound linking is not due to the notorious impermance of onion sites.
    \item If onionsites got drastically more stable, we would still see very low rates of linking.
    \item By elimination of the technological explanation, we suggest that people creating darkweb sites are, on average, simply less social than those creating sites on the www.
\end{enumerate}

The term ``dark web`` is commonplace, but based on our analysis, the ``web'' is a misnomer.  It is more accurate to view it as a set of dark silos.  Unlike the www, the darkweb is a place of isolation.  In \tblref{fig:summary} we summarize our comparison statistics between the www and the darkweb.

\begin{table}[hbt]
\centering

\begin{tabular}{l r r} \toprule
Measure & www \cite{lehmberg2014graph} & darkweb \\
\midrule
    \textbf{Num nodes} & 43M & \numnodes \\
    \textbf{Num edges} & 623M & \numedges \\
    \textbf{Prop. connected pairs} & $\sim\!\! 42\%$ & 8.11\%  \\
    \textbf{Average SPL} & \muSPLwww & \muSPLdw \\
    \textbf{Edges per node} & 14.5 & 3.50 \\
    \textbf{Network diameter}* & 48 & 5 \\
    \textbf{Harmonic diameter} & $\sim\!\!9.55$ & 232.49 \\
\bottomrule
\end{tabular}

\caption{Summarized network level properties between the www and the darkweb.  Asterisk for the entries requiring conversion to an undirected graph.}
\label{fig:summary}
\end{table}

\bibliography{darkweb}

\begin{thebibliography}{10}
\providecommand{\url}[1]{\texttt{#1}}
\providecommand{\urlprefix}{URL }
\expandafter\ifx\csname urlstyle\endcsname\relax
  \providecommand{\doi}[1]{doi:\discretionary{}{}{}#1}\else
  \providecommand{\doi}{doi:\discretionary{}{}{}\begingroup
  \urlstyle{rm}\Url}\fi
\providecommand{\bibAnnoteFile}[1]{%
  \IfFileExists{#1}{\begin{quotation}\noindent\textsc{Key:} #1\\
  \textsc{Annotation:}\ \input{#1}\end{quotation}}{}}
\providecommand{\bibAnnote}[2]{%
  \begin{quotation}\noindent\textsc{Key:} #1\\
  \textsc{Annotation:}\ #2\end{quotation}}
\providecommand{\eprint}[2][]{\url{#2}}

\bibitem{borgatti1998network}
Borgatti SP, Jones C, Everett MG (1998) Network measures of social capital.
\newblock Connections 21: 27--36.
\bibAnnoteFile{borgatti1998network}

\bibitem{wu2006query}
Wu P, Wen JR, Liu H, Ma WY (2006) Query selection techniques for efficient
  crawling of structured web sources.
\newblock In: Data Engineering, 2006. ICDE'06. Proceedings of the 22nd
  International Conference on. IEEE, pp. 47--47.
\bibAnnoteFile{wu2006query}

\bibitem{barabasi1999emergence}
Barab{\'a}si AL, Albert R (1999) Emergence of scaling in random networks.
\newblock science 286: 509--512.
\bibAnnoteFile{barabasi1999emergence}

\bibitem{barabasi2000scale}
Barab{\'a}si AL, Albert R, Jeong H (2000) Scale-free characteristics of random
  networks: the topology of the world-wide web.
\newblock Physica A: Statistical Mechanics and its Applications 281: 69--77.
\bibAnnoteFile{barabasi2000scale}

\bibitem{adamic2000power}
Adamic LA, Huberman BA (2000) Power-law distribution of the world wide web.
\newblock Science 287: 2115--2115.
\bibAnnoteFile{adamic2000power}

\bibitem{kumar2000web}
Kumar R, Raghavan P, Rajagopalan S, Sivakumar D, Tompkins A, et~al. (2000) The
  web as a graph.
\newblock In: Proceedings of the nineteenth ACM SIGMOD-SIGACT-SIGART symposium
  on Principles of database systems. ACM, pp. 1--10.
\bibAnnoteFile{kumar2000web}

\bibitem{albert2001physics}
Albert-L{\'a}szl{\'o}~Barab{\'a}si J (2001) The physics of the web.
\newblock Physics Web http://wwwphysicsweborg/article/world/14/7/09 .
\bibAnnoteFile{albert2001physics}

\bibitem{meusel2015graph}
Meusel R, Vigna S, Lehmberg O, Bizer C (2015) The graph structure in the
  web--analyzed on different aggregation levels.
\newblock The Journal of Web Science 1.
\bibAnnoteFile{meusel2015graph}

\bibitem{rfc7686}
Appelbaum J, Muffet A (2015) The ``.onion'' special-use domain name.
\newblock Internet Engineering Task Force RFC 7686.
\bibAnnoteFile{rfc7686}

\bibitem{everton2012disrupting}
Everton SF (2012) Disrupting dark networks, volume~34.
\newblock Cambridge University Press.
\bibAnnoteFile{everton2012disrupting}

\bibitem{darknet}
De~Domenico M, Arenas A (2017) Modeling structure and resilience of the dark
  network.
\newblock Phys Rev E 95: 022313.
\bibAnnoteFile{darknet}

\bibitem{Serrano2007}
Serrano MA, Maguitman A, Bogu\~{n}\'{a} M, Fortunato S, Vespignani A (2007)
  Decoding the structure of the www: A comparative analysis of web crawls.
\newblock ACM Trans Web 1.
\bibAnnoteFile{Serrano2007}

\bibitem{lehmberg2014graph}
Lehmberg O, Meusel R, Bizer C (2014) Graph structure in the web: aggregated by
  pay-level domain.
\newblock In: Proceedings of the 2014 ACM conference on Web science. ACM, pp.
  119--128.
\bibAnnoteFile{lehmberg2014graph}

\bibitem{tormetrics}
Project TT (2017).
\newblock Tormetrics --- unique .onion addresses.
\newblock
  \urlprefix\url{https://metrics.torproject.org/hidserv-dir-onions-seen.html}.
\bibAnnoteFile{tormetrics}

\bibitem{wiki:torchat}
Wikipedia (2016).
\newblock Torchat --- wikipedia{,} the free encyclopedia.
\newblock
  \urlprefix\url{https://en.wikipedia.org/w/index.php?title=TorChat&oldid=728783759}.
\newblock [Online; accessed 28-February-2017].
\bibAnnoteFile{wiki:torchat}

\bibitem{wiki:tormessenger}
Wikipedia (2017).
\newblock Tor (anonymity network) --- wikipedia{,} the free encyclopedia.
\newblock
  \urlprefix\url{https://en.wikipedia.org/w/index.php?title=Tor_(anonymity_network)#Tor_Messenger&oldid=767770868}.
\newblock [Online; accessed 28-February-2017].
\bibAnnoteFile{wiki:tormessenger}

\bibitem{wiki:ricochet}
Wikipedia (2016).
\newblock Ricochet (software) --- wikipedia{,} the free encyclopedia.
\newblock
  \urlprefix\url{https://en.wikipedia.org/w/index.php?title=Ricochet_(software)&oldid=755044411}.
\newblock [Online; accessed 28-February-2017].
\bibAnnoteFile{wiki:ricochet}

\bibitem{clauset2009power}
Clauset A, Shalizi CR, Newman ME (2009) Power-law distributions in empirical
  data.
\newblock SIAM review 51: 661--703.
\bibAnnoteFile{clauset2009power}

\bibitem{pagerank}
Page L, Brin S, Motwani R, Winograd T (1999) The pagerank citation ranking:
  Bringing order to the web.
\newblock Technical Report 1999-66, Stanford InfoLab.
\newblock \urlprefix\url{http://ilpubs.stanford.edu:8090/422/}.
\newblock Previous number = SIDL-WP-1999-0120.
\bibAnnoteFile{pagerank}

\bibitem{vigna13fibonacci}
Vigna S (2013) Fibonacci binning.
\newblock CoRR abs/1312.3749.
\bibAnnoteFile{vigna13fibonacci}

\bibitem{broder2000graph}
Broder A, Kumar R, Maghoul F, Raghavan P, Rajagopalan S, et~al. (2000) Graph
  structure in the web.
\newblock Computer networks 33: 309--320.
\bibAnnoteFile{broder2000graph}

\bibitem{bollobas2004robustness}
Bollob{\'a}s B, Riordan O (2004) Robustness and vulnerability of scale-free
  random graphs.
\newblock Internet Mathematics 1: 1--35.
\bibAnnoteFile{bollobas2004robustness}

\bibitem{biryukov2013trawling}
Biryukov A, Pustogarov I, Weinmann RP (2013) Trawling for tor hidden services:
  Detection, measurement, deanonymization.
\newblock In: Security and Privacy (SP), 2013 IEEE Symposium on. IEEE, pp.
  80--94.
\bibAnnoteFile{biryukov2013trawling}

\end{thebibliography}

\appendix
\part*{Appendix}

\section{Miscellaneous figures}

\begin{table}[hbt]
\centering
\begin{tabular}{l l r r} \toprule
\textbf{Domain} & \textbf{Title} & \textbf{Out-degree} & \textbf{In-degree} \\
\midrule
	 \texttt{directoryvi6plzm} & Tor Directory: A list of onion & 5582 & 1 \\
	 \texttt{visitorfi5kl7q7i} & VisiTOR Search Engine - Tor Hi & 4367 & 18 \\
	 \texttt{skunksworkedp2cg} & A portal containing lists of l & 2769 & 41 \\
	 \texttt{cratedvnn5z57xhl} & The onion crate REBOOT - Tor h & 2758 & 6 \\
	 \texttt{gxamjbnu7uknahng} & The Uncensored Hidden WikiNew & 848 & 13 \\
\multicolumn{4}{l}{ \rule[\dimexpr.5ex-.2pt]{4pt}{.4pt}\xleaders\hbox{\rule{4pt}{0pt} \rule[\dimexpr.5ex-.2pt]{4pt}{.4pt}}\hfill\kern0pt } \\	 
	 \texttt{torvps7kzis5ujfz} & TorVPS & 498 & 33 \\
	 \texttt{zlal32teyptf4tvi} & Fresh Onions & 478 & 4 \\
	 \texttt{hwikis25cffertqe} & Hidden Wiki & 309 & 15 \\
	 \texttt{w363zoq3ylux5rf5} & Galaxy2 Social network & 294 & 32 \\
	 \texttt{y4yhci7273s2yeqk} & \begin{CJK}{UTF8}{mj}조선위키\end{CJK} & 203 & 9 \\ 
	 \texttt{auutwvpt2zktxwng} & Onion Dir & 186 & 25 \\
	 \texttt{wikiwarixvouhwyn} & Light version of original Hidd & 180 & 7 \\
	 \texttt{hdwikicorldcisiy} & HD Wiki & 168 & 22 \\
	 \texttt{ntcixulmms4275vi} & Hidden Wiki - Outdated and fil & 150 & 6 \\
	 \texttt{soupkso3la22ltl3} & This site contains an onion bl & 146 & 6 \\
	 \texttt{torwikignoueupfm} & TorWiki & 128 & 18 \\
	 \texttt{vxzzqqdt54racf3i} & Tor2Dir: A onion sites list & 119 & 3 \\
	 \texttt{godnotaba36dsabv} & \foreignlanguage{russian}{Годнотаба — мониторинг годноты} & 115 & 5 \\
	 \texttt{pduvohnvbtzph6sg} & pduvohnvbtzph6sg.onion & 103 & 1 \\
	 \texttt{rbaco5flcou46wpd} & Welcome To Dark Web Links \& M & 102 & 5 \\
\bottomrule
\end{tabular}
\caption{The top 15 out-degree hubs on the darkweb.}
\label{tbl:tophubs}
\end{table}


\begin{table}[hbt]
\centering
\begin{tabular}{l l l} \toprule
	\multicolumn{1}{c}{PageRank} & \multicolumn{1}{c}{In-degree} & \multicolumn{1}{c}{Harmonic Centrality} \\
	\midrule
\textbf{Freedom Hosting II} & \textbf{Freedom Hosting II} & \textbf{Freedom Hosting II} \\
\textbf{Blockchain - Bitcoin Bloc} & \textbf{Blockchain - Bitcoin Bloc} & \textbf{Blockchain - Bitcoin Bloc} \\
Outlaw Market & DuckDuckGo - Search engin & \textbf{Tor Ads} \\
\textbf{TorShops} $\dagger$ & \textbf{Grams} $\dagger$ & \textbf{Grams} $\dagger$ \\
TorLinks | .onion Link Li $\dagger$ & SIGAINT & DuckDuckGo - Search engin \\
\textbf{Dream Market - Dark web m} $\dagger$ & TORCH: Tor Search Engine $\dagger$ & SIGAINT \\
\textbf{Tor Ads} & \textbf{Tor Ads} & TORCH: Tor Search Engine $\dagger$ \\
Valhalla Market $\dagger$ & \textbf{TorShops} $\dagger$ & \textbf{Alphabay Market} \\
Silkkitie Market $\dagger$ & main.paraZite \# Anarchy $\dagger$ & main.paraZite \# Anarchy $\dagger$ \\
DreamMarket Forum $\dagger$ & \textbf{Dream Market - Dark web m} $\dagger$ & TorBox - Mailservice \\
TOR FREE SPEECH!ask anyth & \textbf{Alphabay Market} & \textbf{Dream Market - Dark web m} $\dagger$ \\
\textbf{On my website you can upl} $\dagger$ & TorBox - Mailservice & \textbf{TorShops} $\dagger$ \\
\textbf{Alphabay Market} & The Hidden Wiki - Outdate $\dagger$ & OnionWallet \\
\foreignlanguage{russian}{Скрытые Ответы} & \textbf{On my website you can upl} $\dagger$ & USA Citizenship $\dagger$ \\
Hidden Answers pt & OnionWallet & The Hidden Wiki - Outdate $\dagger$ \\
HD Wiki $\dagger$ & HTTP File Server & Example rendezvous points \\
\textbf{Grams} $\dagger$ & The Pirate Bay - Torrent & Apples 4 Bitcoin $\dagger$ \\
DeepDotWeb - Surfacing th $\dagger$ & A portal containing lists $\dagger$ & \textbf{On my website you can upl} $\dagger$ \\
OnionDir - Deep Web Link $\dagger$ & Example rendezvous points & HTTP File Server \\
The Darknet Company & USA Citizenship $\dagger$ & *** Deep Web Radio *** \\
OUTLAW Market & not Evil - Search Tor & EasyCoin bitcoin wallet m \\
Carding Community & *** Deep Web Radio *** & TorLinks | .onion Link Li $\dagger$ \\
	\bottomrule
	\end{tabular}
	\caption{The top 20 sites of the darkweb using PageRank, In-degree, and Harmonic Centrality. Bolded entries are present across all three lists. $\dagger$ entries are in the CORE.}
	\label{tbl:topN}
\end{table}

\begin{table}
\centering
\begin{tabular}{l | l r r r} \toprule
\textbf{Address} & \textbf{Name} & \textbf{Pagerank} & \textbf{In-degree} & \textbf{Out-degree} \\
\midrule
	\texttt{shopsat2dotfotbs} & TorShops & 6.30\textsc{e}-03 & 56 & 1\\
	\texttt{torlinkbgs6aabns} & TorLinks | .onion Link List & 6.00\textsc{e}-03 & 34 & 43\\
	\texttt{lchudifyeqm4ldjj} & Dream Market - Dark web market featuring & 5.59\textsc{e}-03 & 54 & 5\\
	\texttt{valhallaxmn3fydu} & Valhalla Market & 3.75\textsc{e}-03 & 23 & 3\\
	\texttt{silkkitiehdg5mug} & Silkkitie Market & 3.58\textsc{e}-03 & 28 & 7\\
	\texttt{tmskhzavkycdupbr} & DreamMarket Forum & 3.00\textsc{e}-03 & 26 & 6\\
	\texttt{jd6yhuwcivehvdt4} & Dream Market - Dark web market featuring & 2.39\textsc{e}-03 & 23 & 5\\
	\texttt{t3e6ly3uoif4zcw2} & Dream Market - Dark web market featuring & 2.39\textsc{e}-03 & 23 & 5\\
	\texttt{7ep7acrkunzdcw3l} & Dream Market - Dark web market featuring & 2.28\textsc{e}-03 & 17 & 4\\
	\texttt{tt3j2x4k5ycaa5zt} & On my website you can upload/download fi & 1.85\textsc{e}-03 & 42 & 53\\
	\texttt{hdwikicorldcisiy} & HD Wiki & 1.60\textsc{e}-03 & 22 & 168\\
	\texttt{grams7enufi7jmdl} & Grams & 1.60\textsc{e}-03 & 60 & 6\\
	\texttt{deepdot35wvmeyd5} & DeepDotWeb - Surfacing the News From The & 1.57\textsc{e}-03 & 28 & 13\\
	\texttt{dirnxxdraygbifgc} & OnionDir - Deep Web Link Directory & 1.53\textsc{e}-03 & 28 & 65\\
	\texttt{n2ha26oplph454e6} & Welcome To A New Site & 1.28\textsc{e}-03 & 7 & 1\\
	\texttt{rbaco5flcou46wpd} & Welcome To Dark Web Links \& More! & 1.23\textsc{e}-03 & 5 & 102\\
	\texttt{radiocbsi2q27tob} & Rádio CBS – Comunicações Brasileira de S & 9.85\textsc{e}-04 & 8 & 12\\
	\texttt{lwplxqzvmgu43uff} & Runion - Russian Forum & 8.58\textsc{e}-04 & 13 & 5\\
	\texttt{hansamkt2rr6nfg3} & HANSA Market & 8.36\textsc{e}-04 & 34 & 2\\
	\texttt{wallstyizjhkrvmj} & Wall Street Market & 7.99\textsc{e}-04 & 11 & 2\\
	\texttt{w363zoq3ylux5rf5} & Galaxy2 Social network & 7.47\textsc{e}-04 & 32 & 294\\
	\texttt{kpynyvym6xqi7wz2} & main.paraZite \# Anarchy files and Underg & 7.13\textsc{e}-04 & 56 & 71\\
	\texttt{bankshopiweol3mv} & Store & 7.00\textsc{e}-04 & 5 & 2\\
	\texttt{zqktlwi4fecvo6ri} & The Hidden Wiki - Outdated, full of scam & 6.95\textsc{e}-04 & 45 & 145\\
	\texttt{chattorci7bcgygp} & This is ChatTor, the only 100\% anonymous & 6.75\textsc{e}-04 & 16 & 3\\
	\texttt{xmh57jrzrnw6insl} & TORCH: Tor Search Engine & 6.35\textsc{e}-04 & 57 & 9\\
	\texttt{slpwlkryjujyjhct} & SleepWalker & 6.12\textsc{e}-04 & 7 & 2\\
	\texttt{cocaineo5z66elwy} & Concerned Cocaine Citizens & 6.04\textsc{e}-04 & 4 & 5\\
	\texttt{darkmarabrstwfuh} & Darkmarket Market & 5.78\textsc{e}-04 & 7 & 2\\
	\texttt{x7bwsmcore5fmx56} & Darknet Hacking Services & 4.64\textsc{e}-04 & 2 & 1\\
	\texttt{torvps7kzis5ujfz} & TorVPS & 4.49\textsc{e}-04 & 33 & 498\\
	\texttt{wikitorcwogtsifs} & The Hidden Wiki & 4.44\textsc{e}-04 & 11 & 222\\
	\texttt{ntcixulmms4275vi} & Hidden Wiki - Outdated and filled with s & 4.40\textsc{e}-04 & 6 & 150\\
	\texttt{mystorew25hgytln} & Store & 4.35\textsc{e}-04 & 7 & 6\\
	\texttt{fantomwf4luxar7u} & Fantom urls - Forum for paranoids & 3.85\textsc{e}-04 & 4 & 1\\
	\texttt{xfnwyig7olypdq5r} & USA Citizenship & 3.67\textsc{e}-04 & 40 & 1\\
	\texttt{vfqnd6mieccqyiit} & UK Passports & 3.55\textsc{e}-04 & 36 & 1\\
	\texttt{deeplinkdeatbml7} & DeepLink & 3.35\textsc{e}-04 & 4 & 11\\
	\texttt{abbujjh5vqtq77wg} & Onion Identity & 3.35\textsc{e}-04 & 31 & 1\\
	\texttt{54ogum7gwxhtgiya} & Krang Hidden Base in Tor. Technodrome. B & 3.28\textsc{e}-04 & 27 & 18\\
	\texttt{costeirazb2xecgs} & costeira.i2p.onion - Servidor de downloa & 3.21\textsc{e}-04 & 7 & 3\\
	\texttt{allyour4nert7pkh} & AYB -- ur mum XDDDDDDD & 3.18\textsc{e}-04 & 13 & 17\\
	\texttt{linkskgiymtyszdb} & LINKS Onion Web Link Directory - Your co & 3.15\textsc{e}-04 & 10 & 94\\
	\texttt{skunksworkedp2cg} & A portal containing lists of links autom & 3.11\textsc{e}-04 & 41 & 2769\\
	\texttt{tfwdi3izigxllure} & Apples 4 Bitcoin & 3.08\textsc{e}-04 & 39 & 1\\
	\texttt{fdwocbsnity6vzwd} & French Deep Web & 3.06\textsc{e}-04 & 26 & 8\\
	\texttt{roothitpesjylrta} & PT-BR: Site oficial do roothit EN-US: We & 3.02\textsc{e}-04 & 7 & 1\\
	\texttt{tuu66yxvrnn3of7l} & UK Guns and Ammo Store & 3.01\textsc{e}-04 & 31 & 1\\
	\texttt{3dbr5t4pygahedms} & ccpal - ccs - cvv2s - paypal & 3.00\textsc{e}-04 & 31 & 1\\
	\texttt{y3fpieiezy2sin4a} & HQER - High Quality Euro Counterfeits - & 3.00\textsc{e}-04 & 31 & 1\\

\bottomrule
\end{tabular}
\caption{Top 50 domains by Pagerank in the \textsc{CORE}.}
\label{tbl:CORE}
\end{table}

\clearpage
\newpage

\begin{figure}[h!bt]
	\centering
	\includegraphics[width=0.6\textwidth]{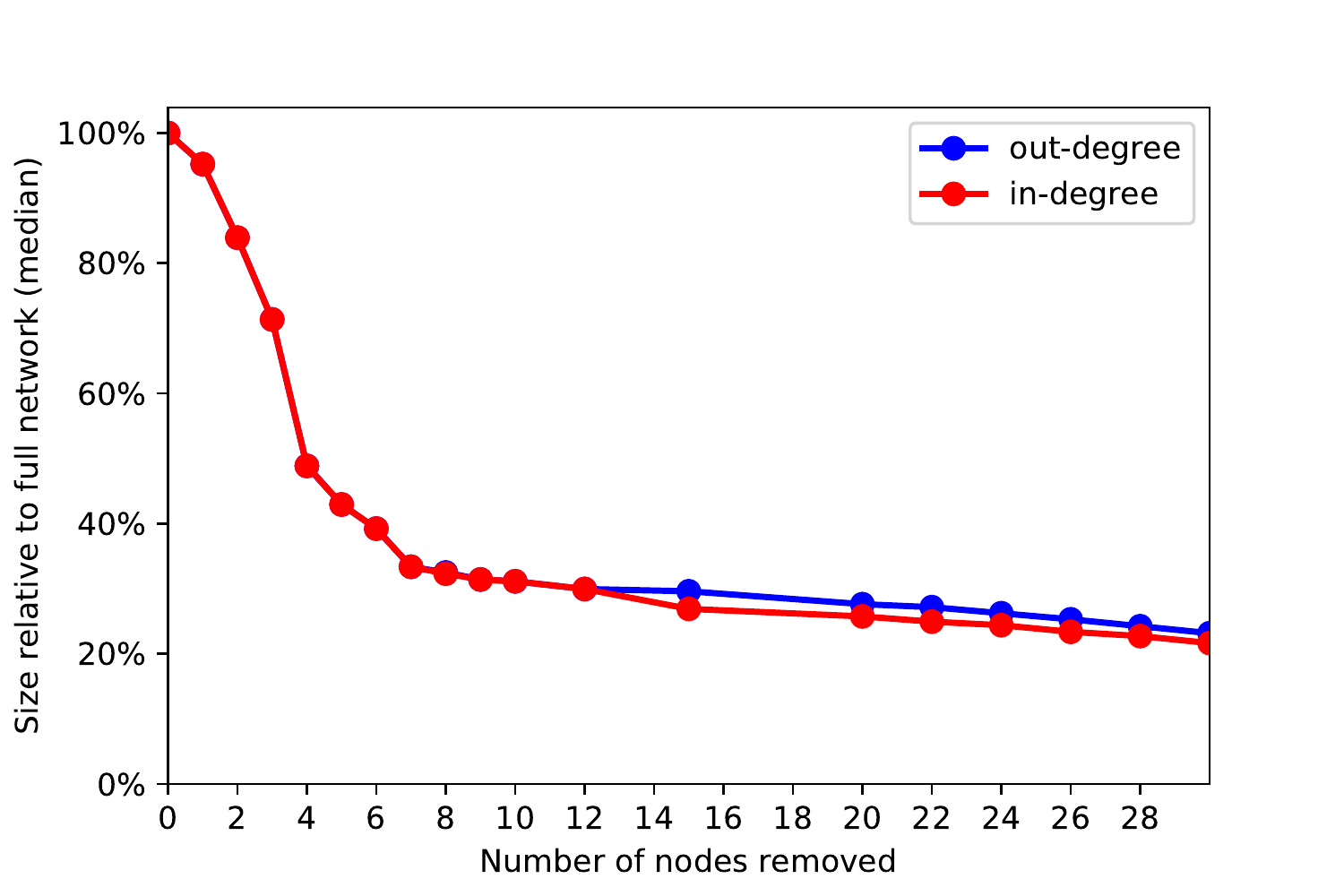}
	\caption{We see that removing the high in-degree and out-degree nodes are about equally effective in breaking the WCC.}
	\label{fig:more_robustness}	
\end{figure}

\section{Peering beneath the Darkweb}
\begin{figure}[h!bt]
	\centering
    \subfloat[Number of introduction points to a darkweb site]{ \includegraphics[width=0.5\textwidth]{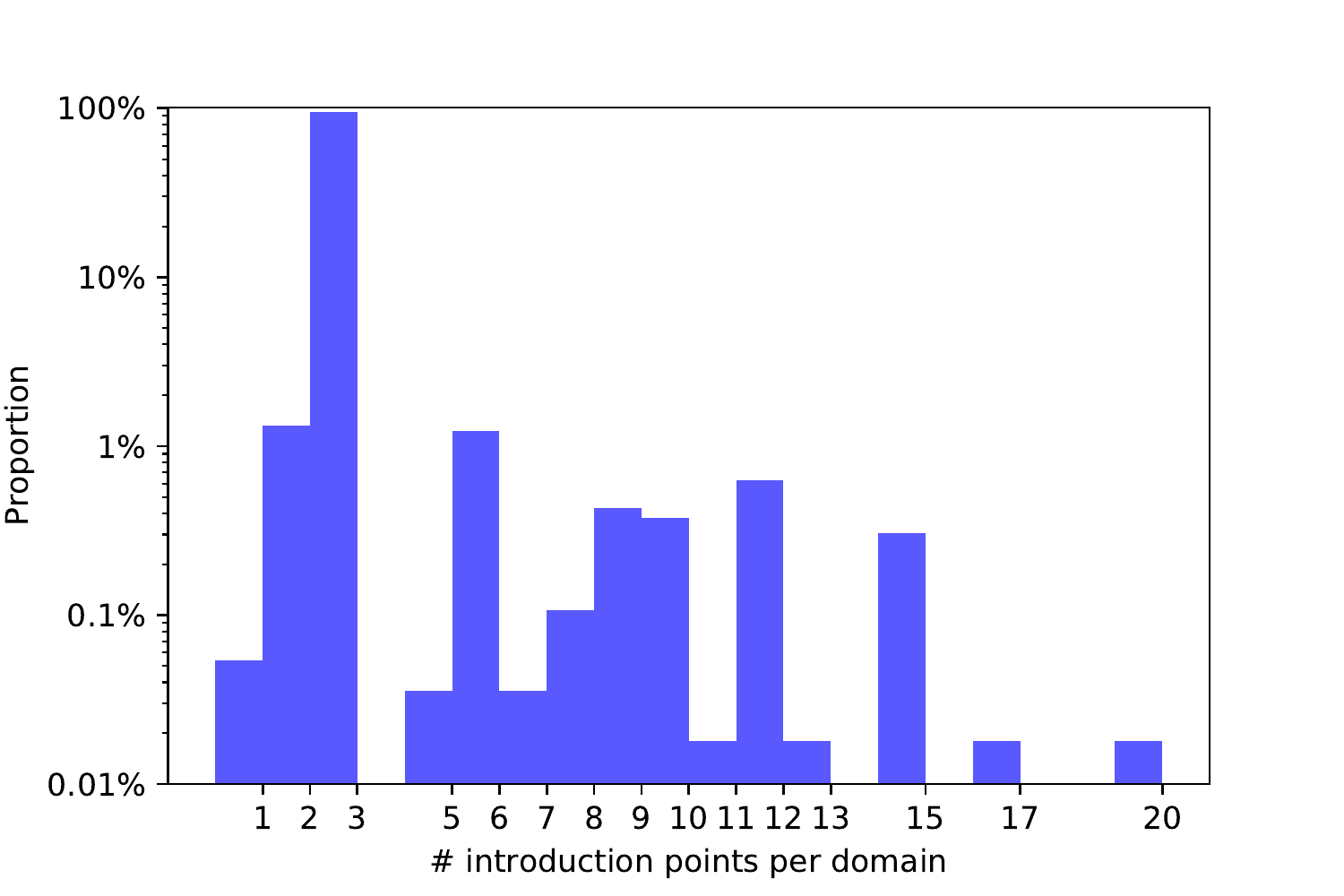} }
	\subfloat[domains per introduction point]{ \includegraphics[width=0.5\textwidth]{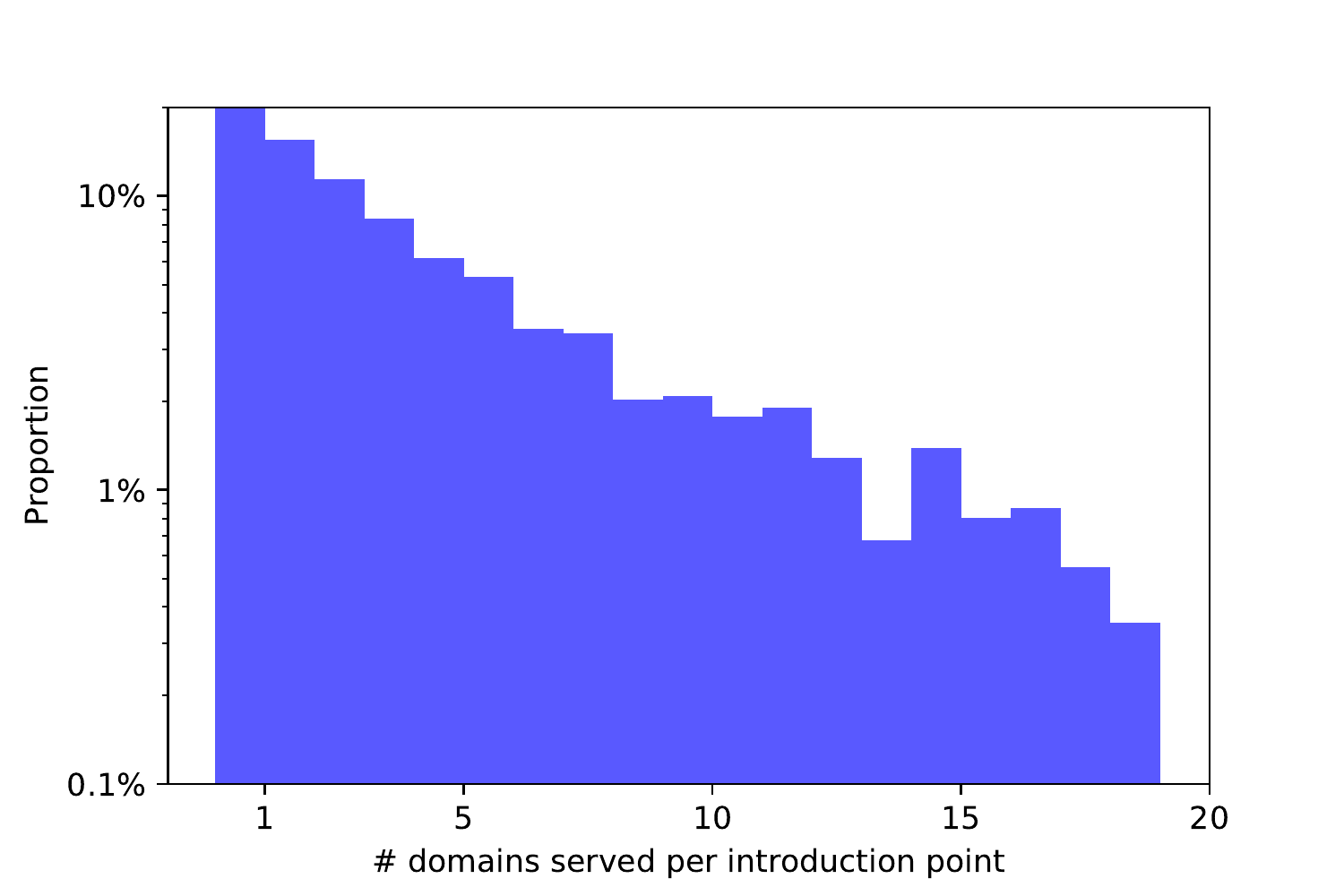} \label{fig:dw_domains_per_intropoints} }


	\caption{Number of domains found on each HSDir}
	\label{fig:dw_domains_per_hsdir}
\end{figure}

Every darkweb domain can be accessed by a site-specified number of ``Introduction Points''.  These introduction points and hidden service directories have been explored before \cite{biryukov2013trawling}, but as far as we know, never statistically.  In our analysis, we found that $96.8\%$ of darkweb sites have exactly $\leq 3$ introduction points (with $95\%$ having exactly $3$).  The most salient fact is that if someone wanted to take down a darkweb site (presumably via DDOS), $97\%$ of the time the attack would require taking out three publicly known relays.  From \ref{fig:dw_domains_per_intropoints}, the number of domains served from each Introduction Point decays roughly exponentially.

\end{document}